 \pdfoutput=1 
\documentclass[
10pt,
]{achemso}

\usepackage{graphicx}

\usepackage{microtype}
\usepackage{booktabs}
\usepackage{soul}
\usepackage[para]{threeparttable}
 \usepackage{lineno}
\usepackage[english]{babel}
\usepackage[utf8]{inputenc}
\usepackage[pdftex]{hyperref, color}
\usepackage[figure, figure*]{hypcap}
\usepackage{amsmath}

\usepackage{upgreek}
\usepackage{lmodern}
\usepackage{wrapfig}
\usepackage{dcolumn}
\usepackage{bm}

\usepackage{bbold}
\usepackage[T1]{fontenc}
\usepackage{color,bm, braket}
\usepackage[emulate = units]{siunitx}

\usepackage{xcolor}

\newcommand {\mo}{MoS$_2$}

\title{Modulated Kondo screening along magnetic mirror twin boundaries in monolayer \mo~on graphene}

\author{Camiel van Efferen}
\email{efferen@ph2.uni-koeln.de}
\affiliation{II. Physikalisches Institut, Universit\"{a}t zu K\"{o}ln, Z\"{u}lpicher Stra\ss e 77, 50937 K\"{o}ln, Germany}

\author{Jeison Fischer} 
\affiliation{II. Physikalisches Institut, Universit\"{a}t zu K\"{o}ln, Z\"{u}lpicher Stra\ss e 77, 50937 K\"{o}ln, Germany}

\author{Theo A. Costi} 
\affiliation{ Peter Gr{\"u}nberg Institut, Forschungszentrum J{\"u}lich, 52425 J{\"u}lich, Germany}
\alsoaffiliation{Institute for Advanced Simulation, Forschungszentrum J{\"u}lich, 52425 J{\"u}lich, Germany}

\author{Achim Rosch}
\affiliation{Institut f\"ur theoretische Physik, Z\"{u}lpicher Stra\ss e 77, 50937 K\"{o}ln, Germany}

\author{Thomas Michely}
\affiliation{II. Physikalisches Institut, Universit\"{a}t zu K\"{o}ln, Z\"{u}lpicher Stra\ss e 77, 50937 K\"{o}ln, Germany}

\author{Wouter Jolie} 
\affiliation{II. Physikalisches Institut, Universit\"{a}t zu K\"{o}ln, Z\"{u}lpicher Stra\ss e 77, 50937 K\"{o}ln, Germany}

\begin{document}
\clearpage

\begin{abstract}
A many-body resonance emerges at the Fermi energy when an electron bath screens the magnetic moment of a half-filled impurity level. This Kondo effect, originally introduced to explain the abnormal resistivity behavior in bulk magnetic alloys~\cite{Kondo1964}, has been realized in many quantum systems over the past decades, such as quantum dots~\cite{Goldhaber1998, Cronenwett1998, Borzenets2020}, quantum point contacts~\cite{Cronenwett2002, Iqbal2013, Rejec2006, Smith2022}, nanowires~\cite{Nygaard2000}, single-molecule transistors~\cite{Liang2002, Park2002, Guo2021}, heavy-fermion lattices~\cite{Jiao2020, Vano2021, Ruan2021}, down to adsorbed single atoms~\cite{Madhavan1998, Li1998d, Trishin2021}.  Here we describe a unique Kondo system which allows us to experimentally resolve the spectral function consisting of impurity levels and Kondo resonance in a large Kondo temperature range, as well as their spatial modulation. Our experimental Kondo system, based on a discrete half-filled quantum confined state within a \mo{} grain boundary, in conjunction with numerical renormalization group calculations, enables us to test the predictive power of the Anderson model which is the basis of the microscopic understanding of Kondo physics.
\end{abstract}

The Kondo effect of single magnetic atoms or molecules on metal surfaces has been the subject of intense research since it was first observed with the scanning tunnelling microscope (STM) by Madhavan et al. \cite{Madhavan1998} and Li et al. \cite{Li1998}. While the Kondo resonance has been well-characterized for numerous such systems~\cite{Ternes2008,Ternes2015}, the underlying atomic or molecular impurity levels that give rise to it have largely remained experimentally inaccessible. This is due to the dominant contribution of substrate states to the tunneling current, which lie in the same energy range as the localized $d$ and $f$ orbitals involved. Furthermore, the strong hybridization of the impurity states with these substrate states, electronvolts away from the Fermi energy, obscures the connection between Kondo resonance and impurity levels. While STM experiments for magnetic atoms on surfaces show that their behaviour close to the Fermi energy can often be understood in terms of the universal physics of 
the Kondo effect, the absence of a full characterization of their impurity levels and Coulomb interactions has led to a strong dependence on theoretical input
to clarify the origin of the measured signals. For example, the line shapes around zero bias, observed in STM spectra for Co on noble metal surfaces~\cite{Madhavan1998} were considered as paradigms for a Kondo resonance, but have recently been argued by Bouaziz et al. to originate from exotic spin excitations rather than the Kondo effect~\cite{Bouaziz2020}. In consequence, theoretical predictions for the dependence of the Kondo resonance on energy position, strength of local interactions, and width of the impurity level \cite{Ternes2015,Schrieffer1966} could not be tested due to the absence of experimental data. 

A promising approach to enable local detection of the impurity levels is to spatially confine electrons, as done in a quantum dot~\cite{Goldhaber1998, Cronenwett1998, Borzenets2020}. When a confined state at the Fermi energy $E_\text{F}$ is filled with a single electron, strong Coulomb repulsion can lift the degeneracy of the energy level, leading to the formation of a singly occupied state below $E_\text{F}$ mimicking the half-filled orbital of a magnetic atom. A Kondo resonance emerges when the non-degenerate states couple to an electron bath in close proximity~\cite{Goldhaber1998, Cronenwett1998, Borzenets2020}. The advantages of the confinement approach are a small separation of the non-degenerate states, together with a large spatial extension of the confined wave function. In quantum dot systems these wavefunctions are, however, not accessible due to the lack of spatial resolution. Here we present a system where scanning tunneling spectroscopy (STS) and STM, with their unsurpassed energy and spatial resolution, can be used to track the Kondo resonance along with the impurity levels and the spatial modulation of their wavefunctions on the atomic scale. It enables us to compare experimental data with unsurpassed precision to predictions for the Anderson model of the Kondo effect obtained through numerical renormalization group (NRG) calculations.

\begin{figure*}
		\centering
		\includegraphics[width=0.9\textwidth]{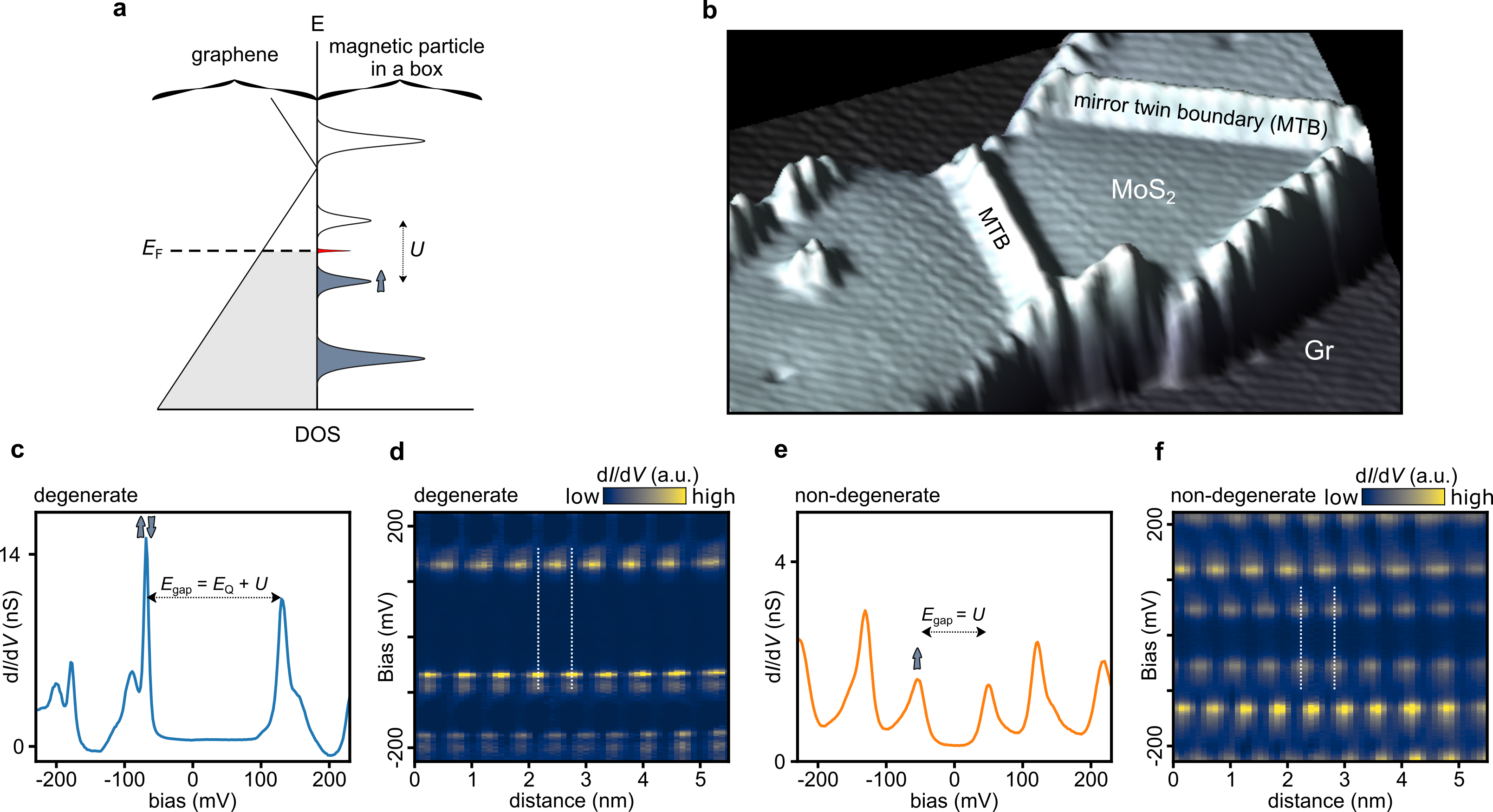}
		\caption{\footnotesize \textbf{Kondo effect within a \mo~MTB. a}~Kondo coupling of graphene (bath) with the non-degenerate states (impurity) confined along a \mo~MTB. Indicated are the Fermi energy $E_\text{F}$ and Coulomb energy $U$, see text. The electron inside the highest occupied states is symbolized by an arrow. \textbf{b}~Atomically-resolved topography of a single-layer \mo{} island on graphene, with two MTBs of lengths $\SI{7.1}{\nm}$ (left) and $\SI{9.0}{\nm}$ (right), separating mirror-symmetric domains. \textbf{c,e}~Spatially averaged d$I$/d$V$ spectra along a MTB of $\SI{8.6}{\nm}$ before \textbf{c} and after \textbf{e} voltage pulses applied with the STM, being in a state that is either \textbf{c} degenerate or \textbf{e} non-degenerate. \textbf{d,f}~Conductance colormaps showing spatially resolved d$I$/d$V$ spectra along the MTB. Dashed white lines are added to highlight the phase relation between the states on either side of $E_\text{F}$ in the middle of the boundary.
		STM/STS parameters: \textbf{b}~\SI[parse-numbers=false]{20 \times 20}{\nano\meter\squared}, $V_\text{set} = \SI{500}{\mV}$, $I_\text{set} = \SI{10}{\pA}$; \textbf{c,d}~$V_\text{set} = \SI{500}{\mV}$, $I_\text{set} = \SI{1.0}{\nA}$; \textbf{e,f}~$V_\text{set} = \SI{500}{\mV}$, $I_\text{set} = \SI{0.5}{\nA}$. $V_\text{mod} = \SI{2.5}{\mV}$.
			\label{Fig1}}
	\end{figure*}

Our Kondo system is realized in a \mo~mirror twin boundary (MTB), a line defect of finite length which hosts confined states in the band gap of the semiconducting two-dimensional (2D) material~\cite{Liu2014dense, Barja2016, Jolie2019a}. Due to its one-dimensional nature, strong Coulomb interactions push the states around the Fermi energy apart and transform higher excitations into the bosonic spin- and charge excitations of a confined Tomonaga-Luttinger liquid~\mbox{\cite{Ma2017angle, Jolie2019a}}. The lowest-energy excitations of such a system can, however, simply be described by a single fermionic level which is either empty, singly occupied, or doubly occupied. Fig.~\ref{Fig1}a sketches the local density of states around the Fermi energy of a MTB placed on a graphene substrate. The electron bath is represented by graphene's Dirac electrons, which exhibit a linear energy dependence close to the Dirac point~\cite{CastroNeto2009b}. The discrete energy levels sketched in Fig.~\ref{Fig1}a, right, are quantized states within the one-dimensional \mo{} MTB. The two energy levels closest to the Fermi energy describe excitations, where a single electron is added to or removed from the MTB. When the highest occupied level is filled by a single electron, the strong Coulomb interaction $U$ prohibits a second electron to enter, creating a spin-$\frac{1}{2}$ system localized along the MTB. This spin couples to the bath and creates through resonant spin-flip processes a Kondo resonance pinned to $E_\text{F}$, as we demonstrate below.

A \mo~monolayer island hosting two such MTBs is shown in Fig.~\ref{Fig1}b. The finite length of the MTBs leads to confined energy levels with a spacing inversely proportional to the length of the wire~\cite{Jolie2019a}. When the highest energy level is filled with two electrons, there is no unpaired magnetic moment and hence no Kondo effect. This situation is depicted in Fig.\ref{Fig1}c, which displays the averaged differential conductance d$I$/d$V$ along a MTB. The d$I$/d$V$ signal of the STM is proportional to the local density of states as a function of energy (given by the bias voltage). We find a series of peaks corresponding to quantized energy levels. Satellite peaks attributed to phonon-induced inelastic tunneling processes are observed at fixed energy intervals $\SI{|24.8| \pm 3.7}{\meV}$ and  $\SI{|47.7| \pm 4.1}{\meV}$ from the main peaks~\cite{Barja2016}. The peaks closest to $E_\text{F}$ exhibit an energy gap $E_\text{gap} = E_\text{Q} + U$, with $E_\text{Q}$~the confinement energy and $U$ the Coulomb energy `penalty' incurred due to strong Coulomb interactions~\cite{Jolie2019a,Yang2021a}. The standing waves corresponding to the energy levels closest to $E_F$ are mapped in Fig.\ref{Fig1}d using a series of d$I$/d$V$ spectra taken along the MTB. These standing waves are out of phase in the center of the boundary, as expected for succesive degenerate particle-in-a-box states (Supplementary Note 1).~\cite{Jolie2019a}. 

The number of electrons within a MTB can be tuned with the help of the STM, either continuously with a back gate or stepwise via voltage pulses in the range $|V_\text{pulse}| = |\SI{1-2.5}|~{\eV}$ (Supplementary Note 2), as has been demonstrated previously for the isostructural 2D material MoSe$_2$~\cite{Yang2021a, Zhu2021}. Fig.~\ref{Fig1}e shows the spectrum obtained on the same boundary after such a voltage pulse. The gap at $E_\text{F}$ is now reduced to the pure Coulomb energy $U$, splitting the formerly degenerate energy level at the Fermi energy. The splitting leads to a singly occupied state below $E_\text{F}$ that is energetically separated from the doubly occupied state by $U$, which in the single electron picture is the unoccupied state visible just above $E_\text{F}$. In the spatially resolved series of d$I$/d$V$ spectra shown in Fig.~\ref{Fig1}f, the non-degenerate nature of the energy levels closest to $E_\text{F}$ is visible as an in-phase beating of the two standing waves above and below the Fermi energy, which would collapse to a degenerate standing wave without Coulomb interaction $U$.

\begin{figure*}
		\centering
		\includegraphics[width=0.9\textwidth]{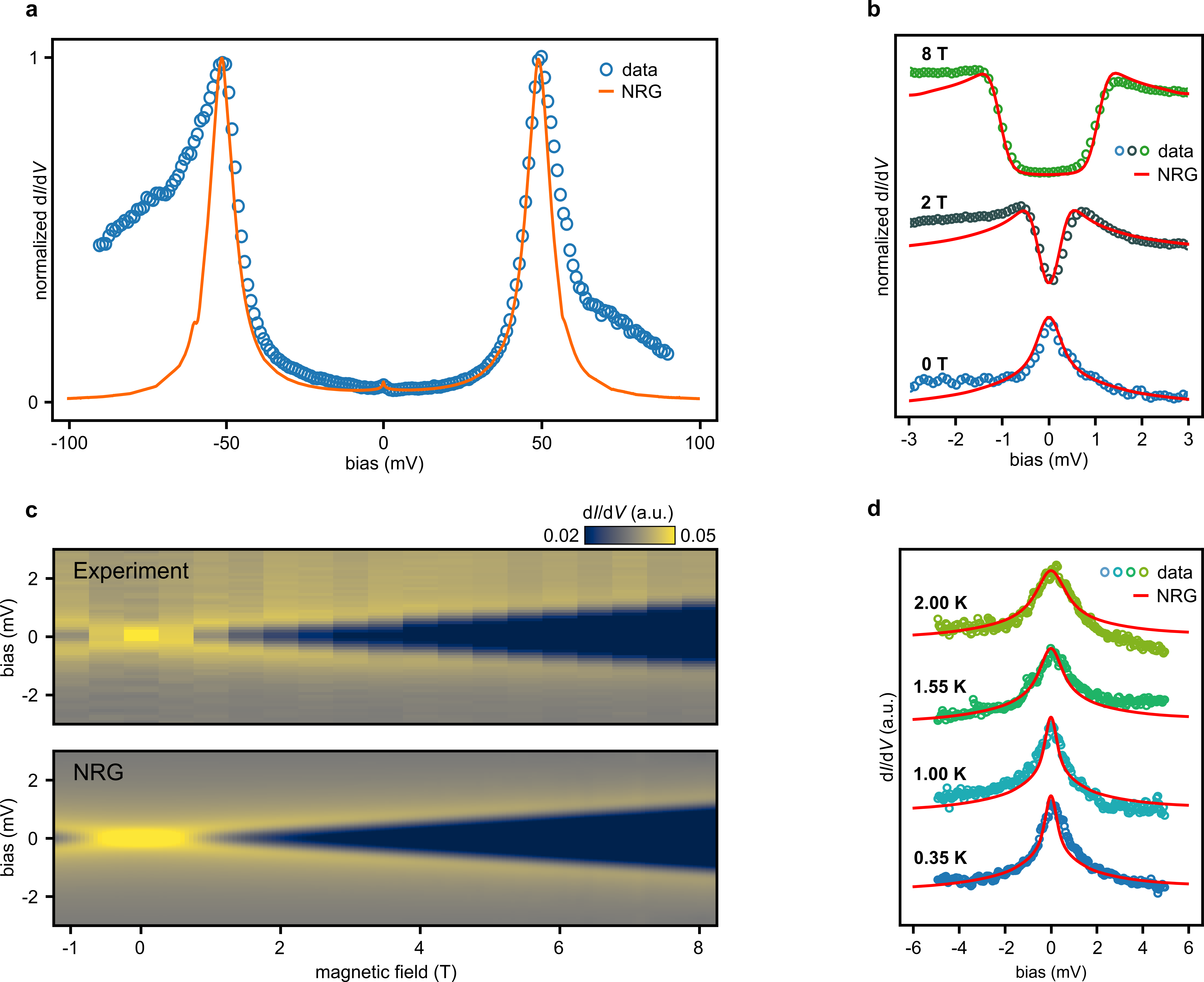}
		\caption{\footnotesize \textbf{Kondo resonance and NRG simulation of magnetic MTB at different magnetic fields and temperatures. a}~d$I$/d$V$ spectrum of impurity level with Kondo resonance (blue circles, $L = \SI{8.6}{\nm}$, $\varepsilon = \SI{-51}{\meV}$, $U = \SI{100}{\meV}$, $\gamma = \SI{10.4}{\meV}$) and corresponding NRG simulation (orange line, $\varepsilon = \SI{-51}{\meV}$, $U = \SI{100}{\meV}$, $\gamma_\text{NRG} = \SI{9.35}{\meV}$). Experimental spectra and theory are normalized to non-degenerate states. The NRG curves have been broadened by the experimental resolution. \textbf{b}~d$I$/d$V$ spectra of Kondo resonance at different magnetic fields (colored circles), with NRG simulation (red line). Plotted in the same scale as \textbf{a}. \textbf{c}~Conductance (d$I$/d$V$ signal) colormaps of the Kondo resonance as a function of bias and magnetic field, comparing experimental spectra (top) and the NRG model (bottom) with $g = 2.5$. \textbf{d}~Dependence of d$I$/d$V$ signal on temperature, with NRG data.
		STM/STS parameters: \textbf{a}~$V_\text{set} = \SI{90}{\mV}$, $I_\text{set} = \SI{1.0}{\nA}$, $V_\text{mod} = \SI{1.0}{\mV}$; \textbf{b,c}~$V_\text{set} = \SI{10}{\mV}$, $I_\text{set} = \SI{1.0}{\nA}$, $V_\text{mod} = \SI{0.2}{\mV}$; \textbf{d}~$V_\text{set} = \SI{5}{\mV}$, $I_\text{set} = \SI{1.0}{\nA}$, $V_\text{mod} = \SI{0.2}{\mV}$
			\label{Fig2}}
	\end{figure*}

In between the two non-degenerate states, a narrow zero-bias peak (ZBP) is found in d$I$/d$V$ spectra, as shown in Fig.~\ref{Fig2}a,b. The ZBP shows a relatively low intensity compared to the non-degenerate peaks (Fig.~\ref{Fig2}a) but becomes well visible when the tip-sample distance is reduced (Fig.~\ref{Fig2}b). No significant change in width and height (relative to the background signal) of the ZBP is observed when varying the tip-sample distance (Supplementary Note 3), while its intensity quickly decays away from the MTB (Supplementary Note 4). To understand the nature of the ZBP, we investigate the influence of magnetic fields and temperature on the shape of the ZBP, shown in Fig.~\ref{Fig2}b-d. We find a clear Zeeman splitting of the ZBP with increasing magnetic field (Fig.~\ref{Fig2}b,c), while the ZBP broadens when the temperature is increased (Fig.~\ref{Fig2}d), in line with the expectation for a Kondo resonance~\cite{Ternes2015}.

		\begin{figure*}
		\centering
		\includegraphics[width=0.9\textwidth]{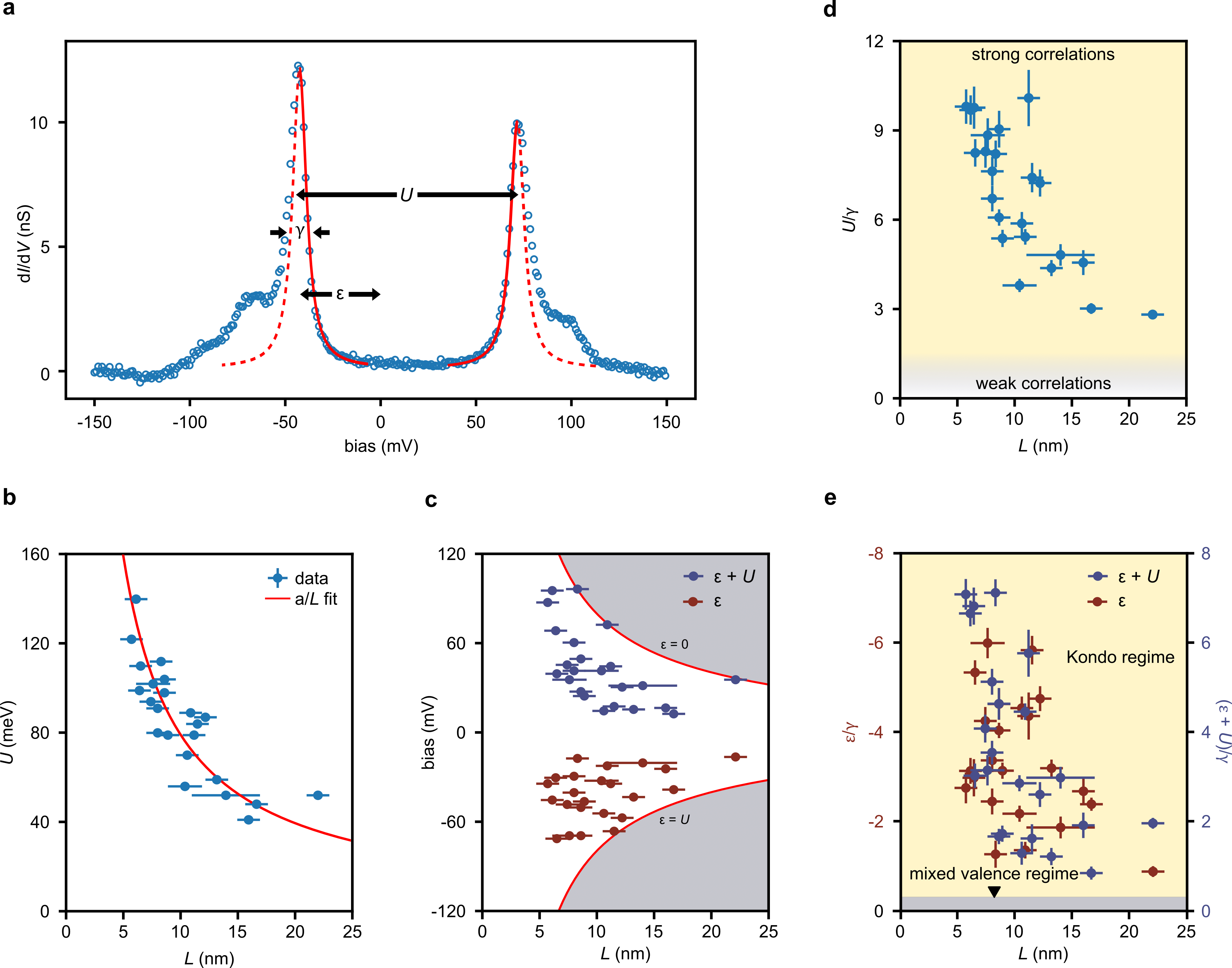}
			\caption{\footnotesize \textbf{Tuning Kondo effect and correlations. a}~d$I$/d$V$ spectrum of non-degenerate states (blue circles) of a MTB with $L = \SI{6.5}{\nm}$, $\varepsilon = \SI{-39}{\meV}$, $U = \SI{110}{\meV}$, $\gamma = \SI{10.8}{\meV}$. Indicated are the Coulomb energy $U$, the energy spacing $\varepsilon$~from $E_\text{F}$, and the full width at half maximum of the main Lorentzian peak $\gamma$, defining the parameters needed for a NRG simulation. Lorentzian functions fitted to the inner slope of the peaks are shown, from which the FWHM $\gamma$ is extracted. \textbf{b}~Coulomb gap ($U$) for MTBs of different length ($L$). $a/L$ fit gives $a = \SI{793}{\meV \nm} = 0.55 e^2/(4 \pi \varepsilon_0)$, where $e$ is the electron charge. \textbf{c}~Peak position of non-degenerate states below ($\varepsilon$) and above ($\varepsilon + U$) the Fermi energy. As $L$ increases, the gap $U$ between the states shrinks. \textbf{d}~Correlation strength $U/\gamma$ of MTBs, calculated for MTBs with $U$ and $\gamma$ obtained from Lorentzian fits. \textbf{e}~Peak position of non-degenerate states in units of $\gamma$. For all boundaries: $|\varepsilon| \geq \gamma/4$ and $|\varepsilon + U| \geq \gamma/4$, placing them in the Kondo regime.
		STM/STS parameters: \textbf{a}~$V_\text{set} = \SI{200}{\mV}$, $I_\text{set} = \SI{0.2}{\nA}$, $V_\text{mod} = \SI{1.0}{\mV}$. The spectra used in \textbf{b, c, d, e} were all measured with $V_\text{mod} = \SI{1.0}{\mV}$, but different stabilization voltages and currents.
			\label{Fig3}}
	\end{figure*}

Theoretically, such a Kondo system is described by the Anderson model, which is established for a single impurity level coupled to an electron bath~\cite{Anderson1970}. The d$I$/d$V$ spectrum of this model is simulated using the numerical renormalization group (NRG) theory and compared to the experimental spectrum in Fig.~\ref{Fig2}a. The NRG simulation resembles the experimental spectrum with striking accuracy for both high and low energy features. This agreement is highlighted when comparing the magnetic field dependence shown for individual spectra in Fig.~2b and as a color plot in Fig.~2c. In both experiment and theory, the resonance transforms into a pronounced gap at high magnetic fields. From the splitting of the Kondo resonance with an out-of-plane magnetic field $B$ we extract a $g$-factor of $g=2.5$. A smaller splitting is observed for in-plane fields (Supplementary Note 5). An excellent match between experiment and theory is also found in the temperature-dependence of the Kondo resonance, depicted in Fig.~2d (Supplementary Note 6). Note that no additional fittting was performed to obtain the spectra in Fig.~2b-d, with the experimental magnetic field and temperature used simply as input for NRG calculations.

The Anderson model parameters required for these simulations are the splitting $U$ of the non-degenerate states, the position $\varepsilon$ of the impurity orbital relative to the Fermi energy and the bare width $\gamma_0$ of the impurity level (Supplementary Note 7). The first two are directly obtained from the experiment, see Fig.~\ref{Fig3}a. A crucial point concerns the extraction of $\gamma_0$ from the measured width $\gamma$. Firstly, as pointed out by by Logan et al.~\cite{Logan1998}, spin flip processes in the Kondo regime increase the experimentally observable width $\gamma$ of the non-degenerate peaks at $\varepsilon$  and $\varepsilon+U$ by a factor of two, $\gamma=2\gamma_0$ (Supplementary Note 8). Secondly, the broad inelastic tail related to phonons leads to a systematic overestimation of the experimental $\gamma$, which we counteract by fitting the inner tail of the non-degenerate peaks (Supplementary Notes 9 \& 11). Taking this into account, we find agreement on a quantitative level between experiment and theory. Importantly, NRG is able to fully predict the magnetic field and temperature dependence of the experimental Kondo resonance purely based on parameters stemming from the impurity level.

Having the capability to predict properties of the Kondo effect using information on the impurity states, we make use of the latter to demonstrate the high tunability of our Kondo system by extracting the parameters ($U,\varepsilon,\gamma$) for 23 different MTBs of varying length $L$. We find a clear inverse length-dependence of $U$ as shown in Fig.~\ref{Fig3}b, which is a consequence of the lower Coulomb energy of electrons when they are spread over a larger $L$ (Supplementary Note 10). A fit $\widetilde{U}(L)=a/L$ is shown as a solid line in Fig.~\ref{Fig3}b. Strong variations are found in $\varepsilon$, which can fluctuate between $-U$~and the Fermi energy. This energy range is visualized in Fig.~\ref{Fig3}c using the fit function $\widetilde{U}(L)$ of Fig.~\ref{Fig3}b. The scatter in $\gamma$, which shows no indication of a length dependence (Supplementary Note 11), is most likely related to the large momentum mismatch of electrons in the MTB and in graphene. This mismatch suppresses direct tunneling and enhances the role of defect-induced tunneling processes.

With the knowledge of ($U,\varepsilon,\gamma$), we first point out that all our non-degenerate states cover a wide range of the strongly correlated regime $U/\gamma\gg 1$, see Fig.~\ref{Fig3}d. Furthermore, all these boundaries satisfy $-\varepsilon/\gamma\gg 1$ and $(\varepsilon+U)/\gamma\gg 1$ showing that these systems are in the Kondo regime, see Fig.~\ref{Fig3}e and Supplementary Note 5. The realization of a wide range of level energies $\varepsilon$ and Coulomb repulsions $U$ in these boundaries translates into a wide range of Kondo temperatures $T_K=10^{-10}\text{ K}-10^{-4}\text{K}$, where $k_BT_K=w\gamma_0\sqrt{U/4\gamma_0}\exp(-\pi|\varepsilon||\varepsilon+U|/\gamma_0U)$ and $w=0.4128$ is the Wilson number (Supplementary Note 12). Such small Kondo temperatures are consistent with the small density of states of the substrate and the large van-der-Waals gap between substrate and MoS$_2$. Since we have access to the microscopic parameters ($U,\varepsilon,\gamma$), $T_\text{K}$ is calculated directly and we do not require estimates of the Kondo coupling to extract $T_\text{K}$. The interested reader will nevertheless find a calculation of Kondo couplings in Supplementary Note 12. These results highlight that there is strong internal tuneablity of the Kondo effect in \mo{} MTBs. Without the need to change the dielectric or chemical environment, a wide range of Kondo temperatures is immediately available due to the wide range of boundary lengths and the asymmetry of the states with respect to the Fermi energy.

			\begin{figure*}
		\centering
		\includegraphics[width=0.45\textwidth]{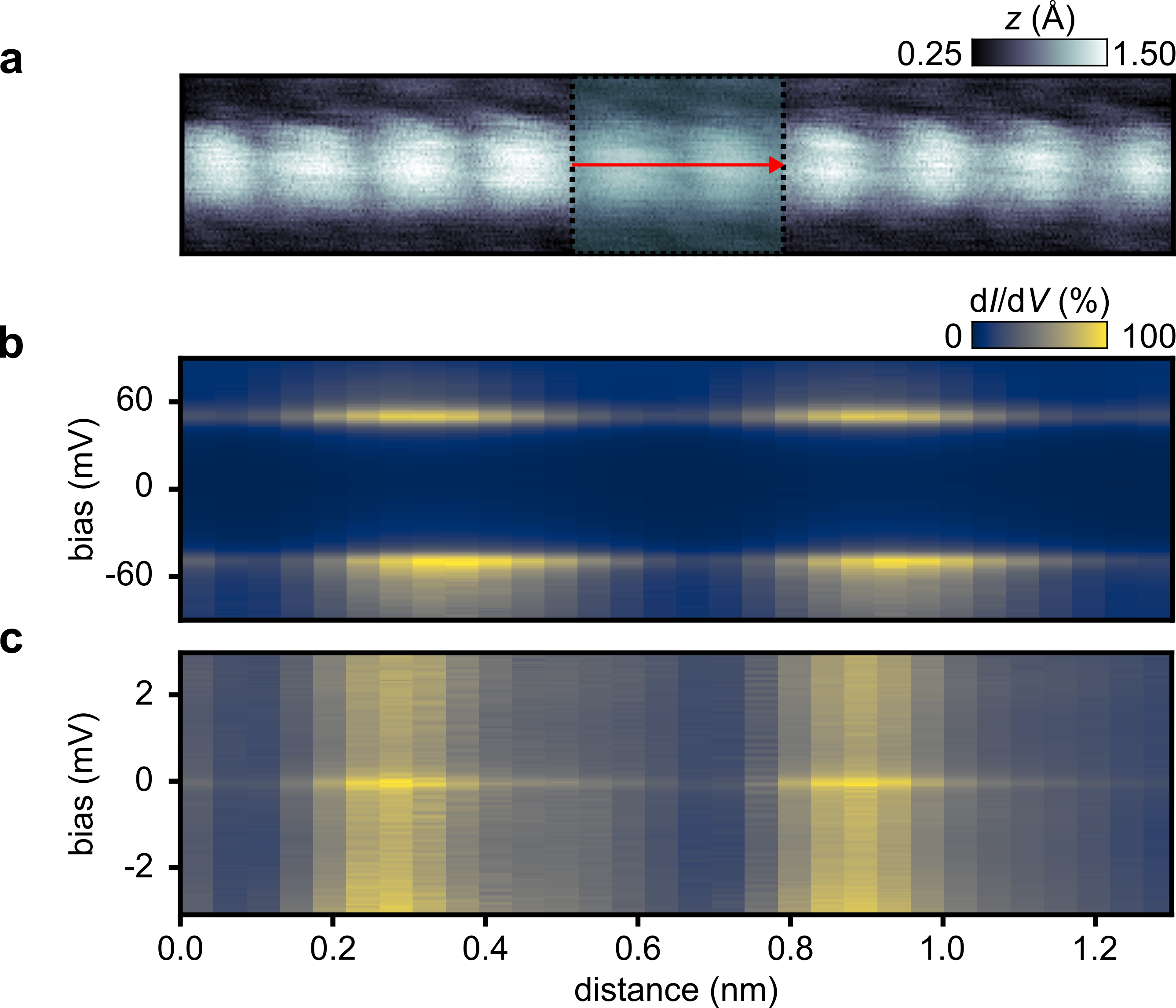}
		\caption{\footnotesize \textbf{Modulated Kondo screening along the particle in a box. a}~Topography along a MTB of $L = \SI{8.6}{\nm}$, $\varepsilon = \SI{-51}{\meV}$, $U = \SI{100}{\meV}$,$\gamma = \SI{9.35}{\meV}$, showing electronic beating of the confined state near $E_\text{f}$. The red arrow indicates where the spectra shown in \textbf{b,c} were taken. \textbf{b}~Conductance colormap of constant height d$I$/d$V$ spectra of the non-degenerate peaks along the MTB. \textbf{c}~Conductance colormap showing constant height d$I$/d$V$ spectra of the Kondo resonance along the same path.
		 STM/STS parameters: \textbf{a}~\SI[parse-numbers=false]{6 \times 1.1}{\nano\meter\squared}, $V_\text{set} = \SI{-10}{\mV}$, $I_\text{set} = \SI{5}{\pA}$; \textbf{b}~$V_\text{set} = \SI{90}{\mV}$, $V_\text{mod} = \SI{1.0}{\mV}$; \textbf{c}~$V_\text{set} = \SI{10}{\mV}$, $V_\text{mod} = \SI{0.2}{\mV}$. $I_\text{set} = \SI{0.5}{\nA}$ for all spectra. 
			\label{Fig4}}
	\end{figure*}
	
Finally, we investigated the spatial distribution of the Kondo resonance together with the non-degenerate states along a MTB. The non-degenerate states beat along the MTB with a wavelength related to their Fermi wave vector \cite{Jolie2019a,Yang2021,Zhu2021}, as expected for a particle in a box. This beating is apparent in the topograph of Fig.~\ref{Fig4}a. Focusing on the two maxima in the boxed area in Fig.~\ref{Fig4}a, the set of d$I$/d$V$ spectra taken along the MTB in Fig.~\ref{Fig4}b shows the beating of the two non-degenerate states on either side of $E_\text{F}$.  The two states are in phase along the MTB as they are derived from the same energy level, which is split due to Coulomb interaction. The d$I$/d$V$ line scan in Fig.~\ref{Fig4}c, taken at the same location, shows the Kondo resonance. The amplitude of the Kondo resonance beats in phase with the non-degenerate peaks, compare Fig.~\ref{Fig4}b. In particular, we find that the amplitude of the resonance is linearly dependent on the peak amplitude of the MTB states, emphasizing the direct relation between these elements of the Kondo effect. These results are reminiscent of the orbital symmetries observed in Shiba states on superconductors which were attributed to anisotropic scattering due to the orbital shapes of the magnetic atom~\cite{Ruby2016, Choi2017}. We envision that using extended magnetic wave functions like those of the MTBs might enable direct access to the correlated behavior of Shiba states and wave function, as observed here for the Kondo effect.

As for the correspondence between Anderson model and experiment, further refining of the model, though outside the scope of this paper, can be envisioned. In particular, as the theoretical model does not include phonons, the inelastic tails and phonon side peaks of the high-energy peaks in Fig.~\ref{Fig3}a are not taken into account. The experimental peaks also vary slightly in height and width, which may arise from the bias dependence of the density of states neglected within our model (the Fermi energy of our graphene layer is estimated to lie \SI{250}{\meV} from the Dirac point~\cite{Ehlen2019}, a substantially larger range than the energy scales relevant for our experiment). Spin-orbit interactions are furthermore expected to induce an anisotropic magnetic response, while multi-channel effects arising from virtual transitions to other high-energy states of the MTB are also neglected in the currrent approach.

In summary, we established 1D magnetic MTBs in a 2D material as a prototypical system to characterize the Kondo effect. We presented a comprehensive study of a fully accessible spin-$\frac{1}{2}$ Kondo system allowing us to not only characterize the Kondo resonance itself, but also to quantitatively determine the impurity levels. We find that NRG calculations accurately reproduce intensity, shape, magnetic field, and temperature dependence of the Kondo resonance when simulating the impurity states. By quantitatively relating high- and low-energy features, our experiment confirms many properties of the Anderson model qualitatively and quantitatively. This includes the ratio of amplitudes of high-and low-energy peaks and an extra broadening of the high-energy peaks by a factor 2 due to spin-flip processes. In addition, our system reveals the real space relation between the impurity level wave function and the spatial dimension of the Kondo resonance. The variability of the parameters $U$ (via boundary length), $\rho$ (via doping~\cite{vanEfferen2022}), and $\varepsilon$ (via pulsing or gating), makes MTBs in transition metal dichalcogenides a valuable testbed for nanoscale magnetic investigations. Controlled coupling of the extended magnetic wave function to localized magnetic defects~\cite{Cochrane2021} or adatoms represent exciting perspectives.

\section*{Methods}
	The \mo~monolayers were grown \textit{in situ} on a graphene substrate, supported by an Ir(111) crystal, in a preparation chamber with base pressure $p < $ \SI{5 \times 10^{-10}}{\milli\bar}. Ir(111) is cleaned by \SI{1.5}{\keV} Ar$^+$ ion erosion and annealing to temperatures $T$ \SI{\approx1550}{\K}. Gr is grown on Ir(111) by two steps. First, room temperature ethylene exposure till saturation followed by \SI{1370}{\K} thermal decomposition gives well-oriented Gr islands. Second, exposure to \SI{2000}{\L} ethylene at \SI{1370}{\K} for \SI{600}{\second} yields a complete single-crystal Gr layer~\cite{Coraux2009}.
	ML \mo{} is grown by Mo deposition in an elemental S pressure of \SI{7 \times 10^{-9}}{\milli\bar}~\cite{Hall2017}.  Subsequently, the sample is annealed to \SI{1050}{\K} in the same S background pressure. 
	
STM and STS are carried out at a base operating temperature of $T_0=$ \SI{0.35}{\K} after \textit{in situ} transfer from the preparation chamber. STS is performed with the lock-in technique, at modulation frequency \SI{907.0}{\Hz}. STM images are taken in constant current mode. Some of the data in Fig.~\ref{Fig3}d was taken using a second STM with an operating temperature of $T=\SI{6.5}{\K}$. 
	
We briefly outline the NRG procedure for the Anderson impurity model. For details, we refer the reader to reviews \cite{Wilson1975, KWW1980a,Bulla2008, Hewson1997}.  
In short, the NRG consists of iteratively diagonalizing the following equivalent linear chain form of the Anderson impurity model (Eq.~(1) in Supplementary Note 5),
\begin{align}
 H=\sum_{\sigma}{\varepsilon_{\sigma}}n_{\sigma}+Un_{\uparrow}n_{\downarrow}+V\sum_{\sigma}(f^{\dagger}_{0\sigma}d_{\sigma}+ \text{H.c.})+\sum_{n=0}^{\infty}\sum_{\sigma}t_n(f_{n\sigma}^{\dagger}f_{n+1\sigma}+\text{H.c.}),\label{eq:AM-linear-chain-form}
\end{align}
where $\varepsilon_{\sigma}=\varepsilon-g\mu_BB\sigma/2$ is  the impurity level energy for spin $\sigma$ measured relative to the Fermi energy, $B$ is a local magnetic field, U is the Coulomb repulsion, and $\gamma_0=2\pi \rho V^2$  is the hybridization strength with $\rho$ the conduction electron density of states. The iterative diagonalization is carried out on a sequence of decreasing energy scales $t_{n}\sim D\Lambda^{-n/2},n=0,1,\dots$, with discretization parameter $\Lambda>1$, by diagonalizing successive truncated Hamiltonians $H_{N}$ with $N=0,1,\dots$ conduction electron orbitals $f_{n\sigma},n=0,\dots,N$, using the recursion relation $H_{N+1}=H_N+ t_{N}\sum_{\sigma}(f_{N\sigma}^{\dagger}f_{N+1\sigma}+\text{ H.c.}) \equiv {\mathcal T}[H_N]$. This procedure yields the many-body eigenvalues and eigenvectors and also the matrix elements of physical observables of interest. The logarithmic discretization parameter
$\Lambda>1$ separates out the many (infinite) energy scales of the conduction band, from high energies (small $n$) to intermediate energies (intermediate $n$) and low energies ($n \gg 1$), allowing the
physics to be obtained iteratively on each successive energy scale. This nonperturbative approach allows essentially exact calculations for
thermodynamical and dynamical quantities, including the spectral function, to be carried out on all temperature and energy scales. In fitting the calculated d$I$/d$V$ to experiment, a constant elastic background of less than $1\%$ of the total intensity takes into account direct tunneling from the substrate.

In order to compare NRG with experiment (see also Supplementary Note 13), it is important to take the experimental broadening due to temperature and lock-in modulation into account. Experimentally, we use an oscillating voltage with an amplitude of $V_\text{mod}=\SI{0.2}{\meV}$, $V(t)=V_0+V_\text{mod} \cos(\omega t)$. As the experimental differential conductance (proportional to $A(\omega)$) is obtained from the standard lock-in technique, we obtain a spectral function $ A(\omega)$ broadened by the experimental resolution from  $\tilde A(V_0)=\frac{1}{2 T V_\text{mod}} \int_0^T dt \cos(\omega t) \int_0^{V(t)} A(\omega)d\omega$. Concerning the temperature implemented in NRG, we use the experimental temperature. At or below $\SI{0.7}{\K}$, a constant $T_{\rm{eff}}=0.7$~K is implemented, which is equal to the temperature of the tunnel junction measured at the lowest experimental temperatures \cite{Bagchi2022}.

\section*{Data availability}
Data that support the plots within this paper and other findings of this study are available from the corresponding author upon reasonable request.

\section*{Acknowledgments}
This work was funded by the Deutsche Forschungsgemeinschaft (DFG, German Research Foundation) - Project number 277146847 - CRC 1238 (subprojects A01, B06 and C02). JF acknowledges financial support from the DFG SPP 2137 (Project FI 2624/1-1) T.A.C. gratefully acknowledges the computing time granted through JARA on the supercomputer JURECA at Forschungszentrum Jülich.

\section*{Author contributions}
W. J. designed the experiments. C. v. E., J. F. and W. J. carried out the measurements. C. v. E. did most of the analysis of the experimental data under the supervision of W. J. T. A. C. performed the NRG calculations and T.A.C and A.R. both gave theoretical input. C. v. E., W. J. and T.M. mainly wrote the manuscript, with help from all the authors. All authors contributed to the discussion and interpretation of the results.

\section*{Ethics declarations}
The authors declare no competing interests.

	\bibliographystyle{naturemag}
	\bibliography{./library}

\end{document}


\clearpage

\newpage

\tableofcontents

\newpage

\section*{Supplementary Note 1: d$I$/d$V$ maps of the confined states close to the Fermi energy}
    \addcontentsline{toc}{section}{Supplementary Note 1: d$I$/d$V$ maps of the confined states close to the Fermi energy}

\begin{figure*}[t!]
		\centering
		\includegraphics[width=\textwidth]{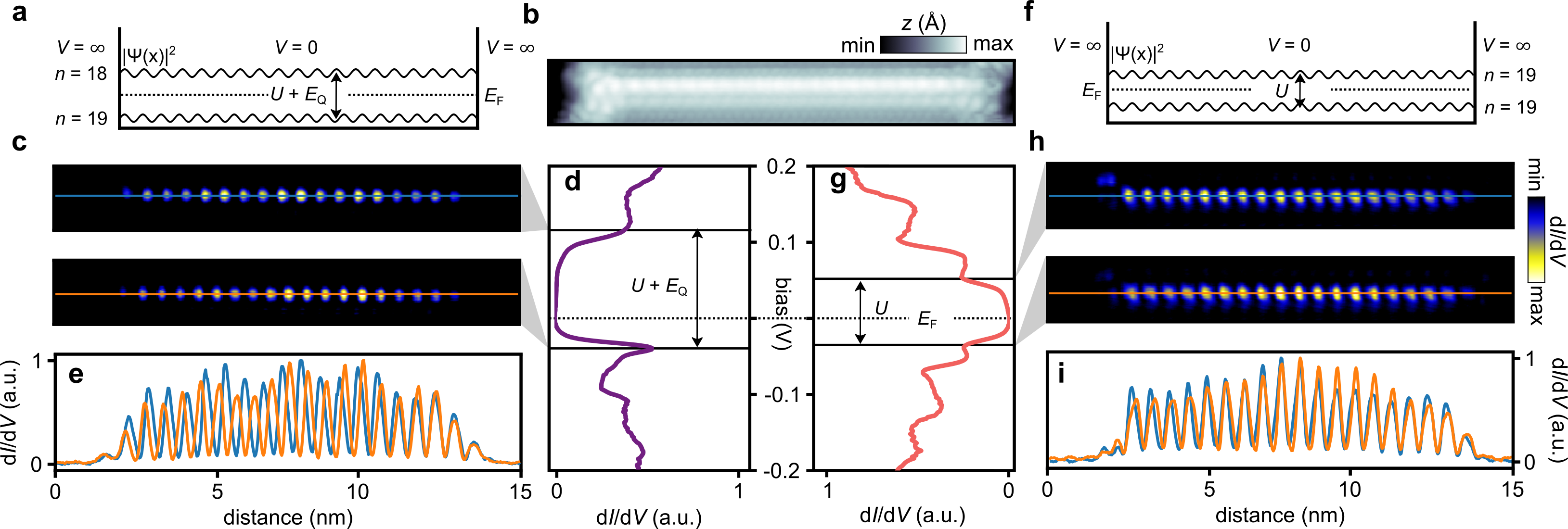}
		\caption{\footnotesize \textbf{Mapping the confined wave functions in mirror twin boundaries. a}~Sketch of the two confined states next to the Fermi energy in a spin-degenerate state.
		\textbf{b}~STM image of the mirror twin boundary.
		\textbf{c}~Maps showing the d$I$/d$V$ signal at the energy of the lowest unoccupied and highest occupied state along the entire length of the boundary. The number of maxima of the wave trains differ by one.
		\textbf{d}~Corresponding d$I$/d$V$ spectra averaged over the length of the boundary.
		\textbf{e}~Linescans along the lines sketched in c, showing the anti-phase behavior of the two wave trains in the center of the boundary.
		\textbf{f}~Sketch of the two states next to the Fermi energy when the highest occupied state is occupied by a single electron.
		\textbf{g}~Spectra averaged over the length of the boundary.
		\textbf{h}~Maps showing the d$I$/d$V$ signal at the energy of the lowest unoccupied and highest occupied state along the entire length of the boundary. Both wave trains have the same number of maxima.
		\textbf{h}~Linescans along the lines sketched in \textbf{c}, showing the in-phase behavior of the two wave trains along the entire boundary. 
		STM/STS parameters: \textbf{b}~\SI[parse-numbers=false]{15 \times 2.1}{\nano\meter\squared}, $V_\text{set} = \SI{1.0}{\V}$, $I_\text{set} = \SI{5}{\pA}$; \textbf{c}~$V_\text{set} = \SI{110}{\mV}$ (top), $V_\text{set} = \SI{-42}{\mV}$ (bottom); \textbf{d,g}~$V_\text{set} = \SI{200}{\mV}$, $I_\text{set} = \SI{0.2}{\nA}$; \textbf{h}~$V_\text{set} = \SI{60}{\mV}$ (top), $V_\text{set} = \SI{-40}{\mV}$ (bottom);  $V_\text{mod} = \SI{2.0}{\mV}$, $f_\text{mod} = \SI{833.0}{\Hz}$ for all spectra.
			\label{SFigfull}}
	\end{figure*}

Supplementary Figure~\ref{SFigfull}~summarizes how we can experimentally distinguish a singly occupied, non-degenerate confined state from a doubly occupied degenerate state. Let us start by pointing out that mirror twin boundaries in \mo{} host a confined Tomonaga-Luttinger liquid that displays spin and charge excitations. However, the first confined states below and above the Fermi energy, the so-called zero modes, can be viewed as single electronic levels which are either empty, singly occupied or doubly occupied. Thus, spin-charge separation does not enter the Kondo problem and consequently it does not complicate the analysis, as long as one focuses on the zero modes only.

The probability density $|\Psi(x)|^2$ of the confined zero modes is what one expects for an electron in a one-dimensional box, \textit{i.e.} it is proportional to $\sin(x)^2$ with the boundary condition of vanishing probability density at the box ends (where we assume a potential energy barrier $V=\infty$). Supplementary Figure~\ref{SFigfull}a represents a sketch for the case of two wave trains with 18 and 19 maxima. In contrast to the non-interacting particle-in-a-box situation, these two degenerate levels are spaced by an energy $E_\text{Q}+U$ , where  $E_\text{Q}$ is the level spacing one would expect for non-interacting electrons and $U$ is the Coulomb penalty when an additional electron is added to the system.

Supplementary Figure~\ref{SFigfull}b shows an atomically resolved STM image of a mirror twin boundary with its highest occupied state being spin-degenerate. Maps of the d$I$/d$V$ signal at the energies of the first quantized states above and below the Fermi energy are presented in Supplementary Figure~\ref{SFigfull}c. Their location in energy is deduced from d$I$/d$V$ spectra measured along the boundary, see Supplementary Figure~\ref{SFigfull}d. The wave trains differ by one maximum (18 at $V = \SI{110}{\mV}$ and 19 at $V = \SI{-42}{\mV}$], best seen in the line profiles taken along the line sketched in the d$I$/d$V$ maps, shown in Supplementary Figure~\ref{SFigfull}e. This difference leads to an out-of-phase behavior in the center of the boundary. Note that the number of maxima increases with decreasing energy due to the hole-like (rather than electron-like) band of the mirror twin boundary~\cite{Jolie2019a}. The excellent agreement with ab initio density functional theory calculations~\cite{Jolie2019a}, together with the ability to shift the periodicity of the states at the Fermi energy by changing the chemical potential~\cite{vanEfferen2022}, further validates our assignment that these wave trains are confined states and have no other structural origin.

Next, we pulse this mirror twin boundary into the non-degenerate state, in which the highest occupied state is filled by one electron at $\varepsilon$, while the lowest energy state above the Fermi energy corresponds to the same quantized state filled with two electrons at $\varepsilon+U$. The sketch in Supplementary Figure~\ref{SFigfull}f visualizes the wave trains above and below the Fermi energy, which have the same number of maxima (19).

After extracting the energy of the quantized states above and below the Fermi energy using d$I$/d$V$ spectra measured along the boundary (Supplementary Figure~\ref{SFigfull}g), we map the two wave trains and find the same number of maxima (19), see Supplementary Figure~\ref{SFigfull}h. The corresponding line profiles in Supplementary Figure~\ref{SFigfull}i lie on top of each other, as the wave trains would collapse to a spin-degenerate level for a vanishing $U$.

\newpage

\section*{Supplementary Note 2: Tuning mirror twin boundary filling with voltage pulses}
    \addcontentsline{toc}{section}{Supplementary Note 2: Tuning mirror twin boundary filling with voltage pulses}

	\begin{figure*}[h!]
		\centering
		\includegraphics[width=0.7\textwidth]{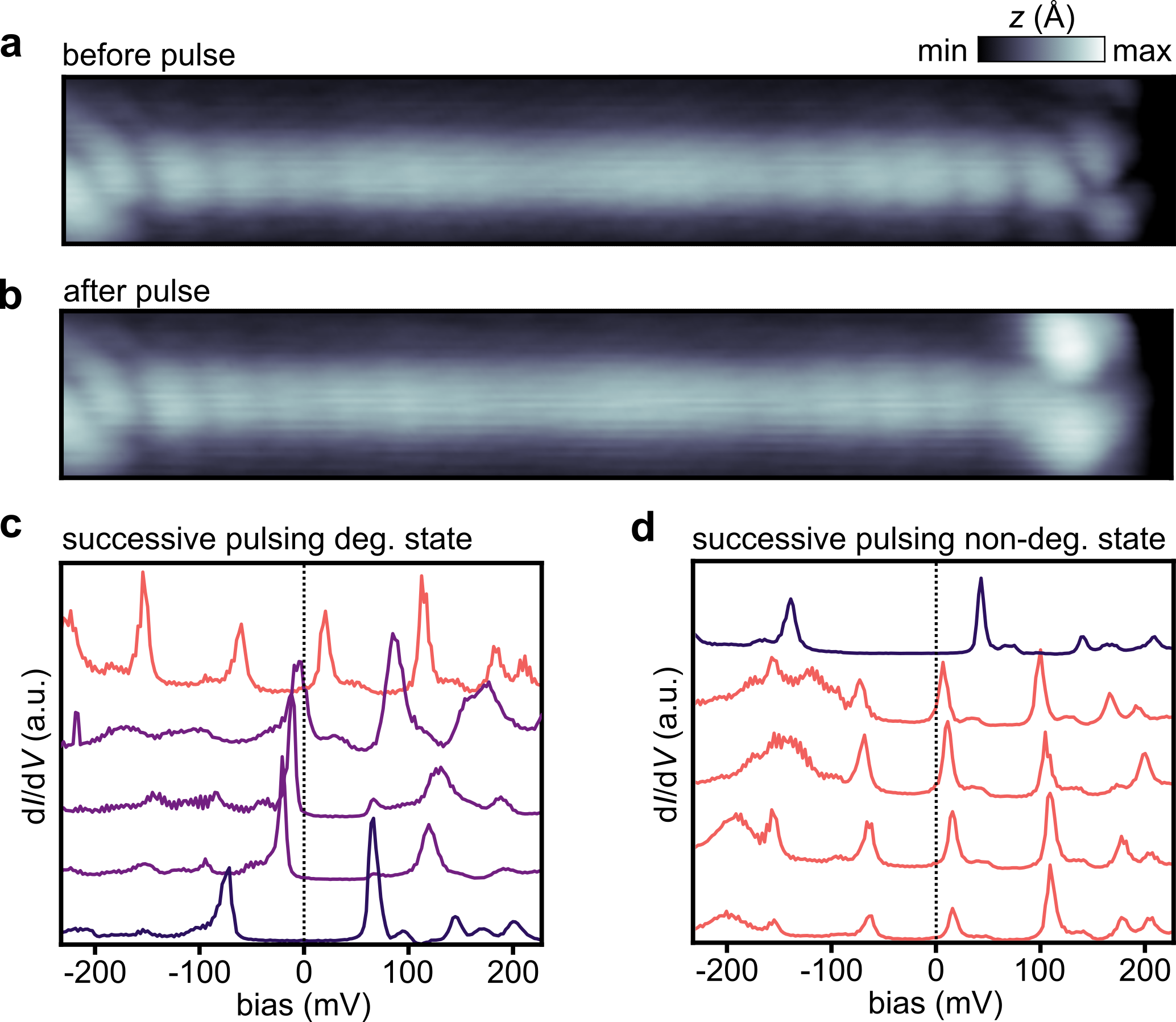}
		\caption{\footnotesize \textbf{Tuning mirror twin boundary filling with voltage pulses. a}~STM topograph of a mirror twin boundary before and \textbf{b}~after the application of a voltage pulse at the right side of the boundary. \textbf{c}~d$I$/d$V$ spectra taken at the same location on a mirror twin boundary after successive voltage pulses. The highest occupied state transforms from a degenerate (bottom) to a non-degenerate state (top). \textbf{d}~Sequence of d$I$/d$V$ spectra taken at the same location on a mirror twin boundary. The highest occupied state transform from a non-degenerate (bottom) to a degenerate state (top). Between each spectrum from bottom to top a voltage pulse in the range $V = \SI{|1 - 2.5|}{\eV}$ was applied.  
		STM/STS parameters: \textbf{a,b}~\SI[parse-numbers=false]{12.0 \times 1.8}{\nano\meter\squared}, $V_\text{set} = \SI{500}{\mV}$, $I_\text{set} = \SI{10}{\pA}$; \textbf{c}~$V_\text{set} = \SI{200}{\mV}$, $I_\text{set} = \SI{500}{\pA}$; \textbf{d}~$V_\text{set} = \SI{200}{\mV}$, $I_\text{set} = \SI{350}{\mA}$. $V_\text{mod} = \SI{2.5}{\mV}$, $f_\text{mod} = \SI{907.0}{\Hz}$ for all spectra.
			\label{SFigPulse}}
	\end{figure*}

Following the method set out in Ref.~\citenum{Yang2021a}, we change the electronic configuration of mirror twin boundaries \textit{via} the application of voltage pulses in the range $V = \SI{|1 - 2.5|}{\eV}$. We find that pulses are most effective near the ends of the mirror twin boundaries, close to the edges of the \mo~islands. Supplementary Figure~\ref{SFigPulse}a,b shows the same mirror twin boundary before and after a voltage pulse performed near the right end of the boundary. Comparing the two STM images, we find that the apparent height at the edge is altered close to the location of the mirror twin boundary. This behavior is reproducibly found after pulsing and fully reversable. While we are not able to provide a microscopic model of the structure modifications at the edge of \mo~islands, the edge modifications induced by voltage pulses are likely related to the shifts in peak position in the mirror twin boundaries observed in d$I$/d$V$.

An example is shown in Supplementary Figure~\ref{SFigPulse}c, where successive pulses on a degenerate boundary (dark purple line) lead first to a shift of the quantized levels such that a single state crosses the Fermi energy (purple lines) and ultimately a transformation to a non-degenerate state (orange line). Note that when a single state overlaps with the Fermi energy, the MTB is in the so-called mixed valence regime and can no longer be regarderd as a pure Kondo system (see Supplementary Note 5 for more information). Conversely, pulsing a non-degenerate boundary, as shown in Supplementary Figure~\ref{SFigPulse}d, can transform the lowest occupied state from non-degenerate (orange lines) into a degenerate state (dark purple line).

\newpage

\section*{Supplementary Note 3: Influence of the tip-sample distance on the Kondo resonance}
    \addcontentsline{toc}{section}{Supplementary Note 3: Influence of the tip-sample distance on the Kondo resonance}

	\begin{figure*}[h!]
		\centering
		\includegraphics[width=0.6\textwidth]{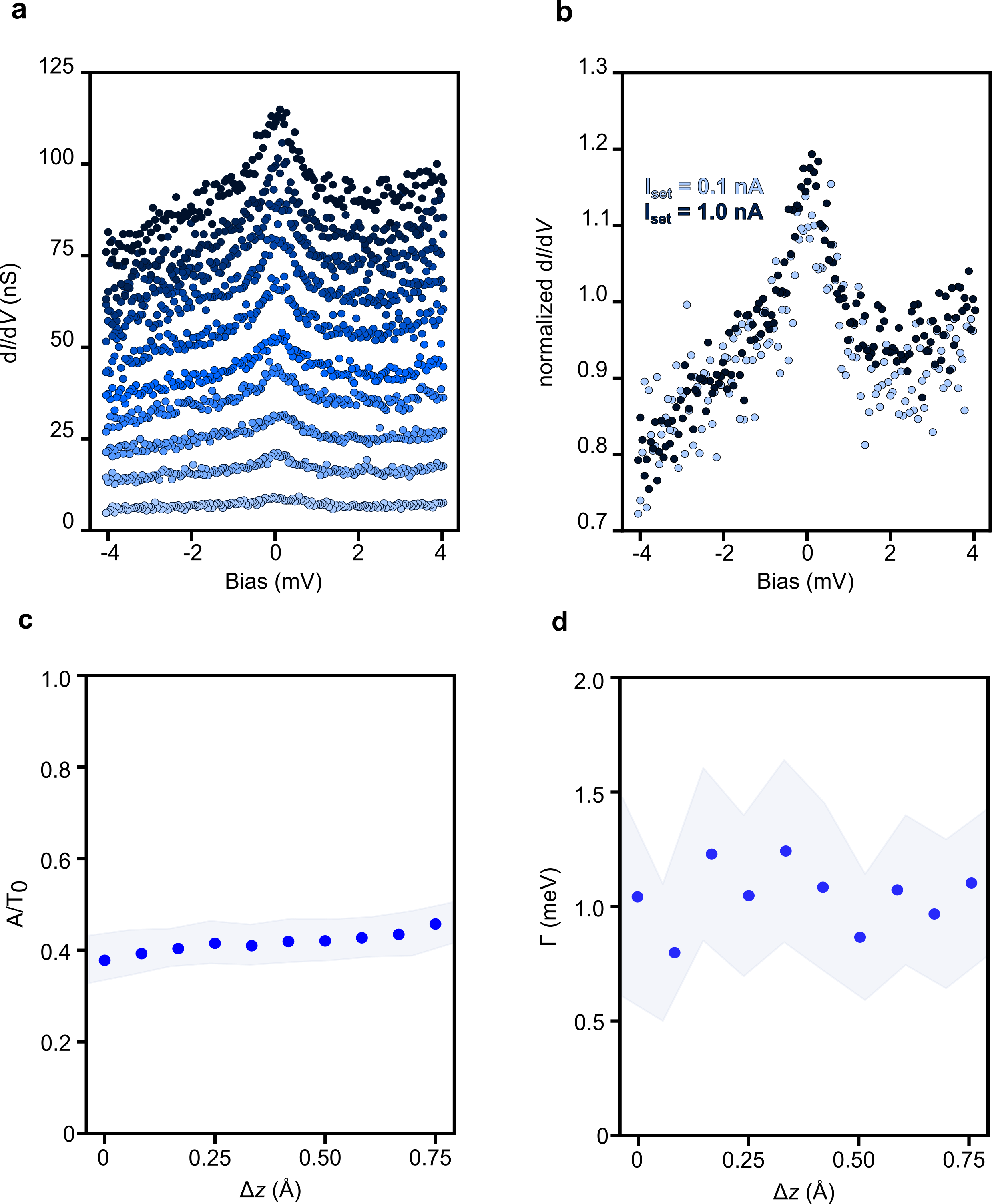}
		\caption{\footnotesize \textbf{Tip-sample distance dependence of Kondo resonance. a}~d$I$/d$V$ spectra obtained with various stabilization currents averaged over the same location. The lowest stabilization current is $\SI{0.1}{\nA}$, and increases stepwise by $\SI{0.1}{\nA}$ between measurements. \textbf{b}~d$I$/d$V$ spectra obtained with $I_\text{set} (\SI{10}{\mV}) = \SI{0.1}{\nA}$ (blue line) and $I_\text{set} (\SI{10}{\mV}) =\SI{1.0}{\nA}$ (green line) stabilization current, normalized so that d$I$/d$V$ ($\SI{4}{\mV}$) $= 1$. \textbf{c, d}~Normalized amplitude A/T$_0$ ($\Delta z$) (c) and width $\Gamma$ (d) of Frota fits (blue dots), with a $2\times$ standard deviation confidence interval indicated in gray. $V_\text{mod} = \SI{0.2}{\mV}$, $f_\text{mod} = \SI{907.0}{\Hz}$.
			\label{SFigCD}}
	\end{figure*}

To rule out that the metallic tip apex of the scanning tunneling microscope (STM) is involved in the formation of the Kondo resonance, we measured a series of d$I$/d$V$ spectra with progressively higher stabilization currents $I_\text{set}$ (\SI{10}{\mV}) $= \SI{0.1 - 1.0}{\nA}$, shown in Supplementary Figure~\ref{SFigCD}a. 

If the electrons in the tip would contribute to Kondo screening, this effect should be enhanced when the tip is brought in close proximity to the non-degenerate states, leading to an increase in the width of the resonance peak. We find however, that an increase in the tunneling current by a factor 10, achieved by bringing the tip $\approx \SI{0.75}{\AA}$ closer to the surface, has only a minor influence on the spectral shape of the Kondo resonance, which is depicted in Supplementary Figure~\ref{SFigCD}b. To quantify the dependence on tip-sample distance, we fitted the spectra with a Frota function to obtain the peak amplitude $A/T_0$ and width $\Gamma$. In Supplementary Figure~\ref{SFigCD}c it can be seen that the amplitude slightly increases by about $10\%$, whereas no detectable change is observed in $\Gamma$ within the confidence interval of the fit.

\newpage

\section*{Supplementary Note 4: Decay of Kondo resonance away from MTB}
    \addcontentsline{toc}{section}{Supplementary Note 4: Decay of Kondo resonance away from MTBs}
\begin{figure*}[h!]
		\centering
		\includegraphics[width=0.9\textwidth]{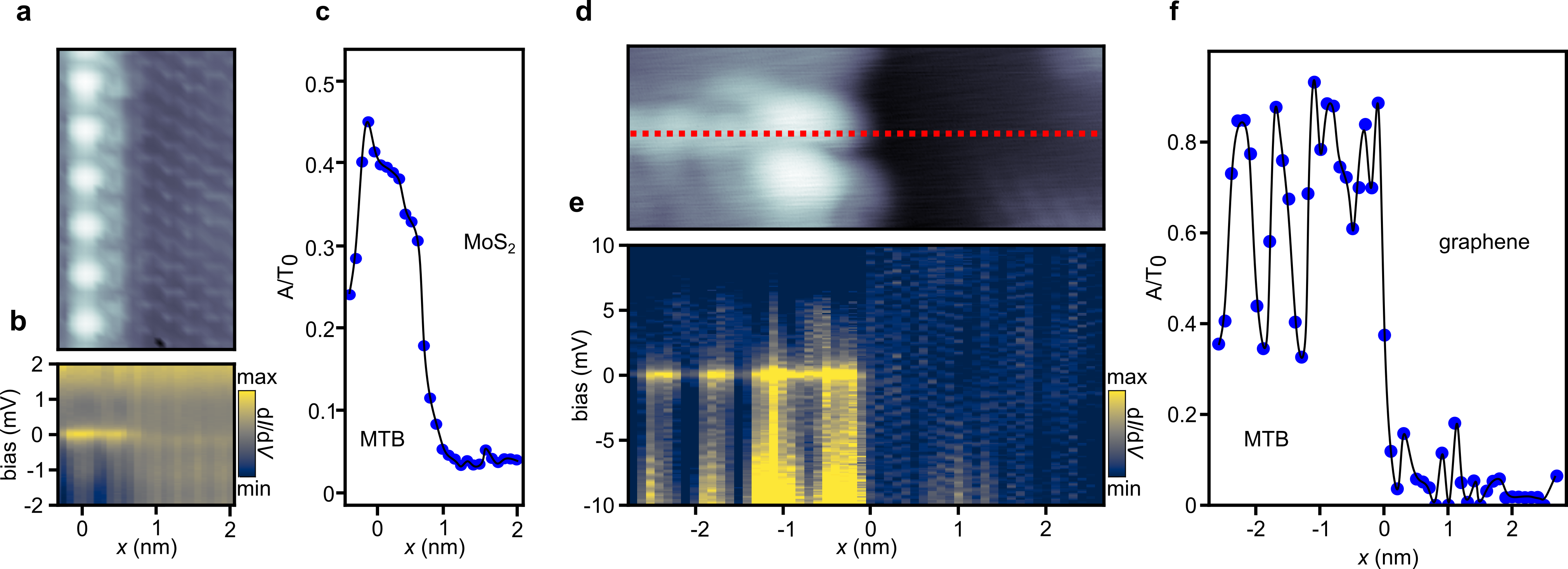}
		\caption{\footnotesize \textbf{Decay of Kondo resonance away from mirror twin boundaries. a}~Topography of \mo{} close to a mirror twin boundary (MTB). \textbf{b}~d$I$/d$V$ spectra averaged along the length of the boundary and \textbf{c}~normalized amplitude A/T$_0$ of Frota fit as a function of sample bias and perpendicular distance $x$ (nm) from center of MTB. \textbf{d}~Topography of \mo{} island edge. Part of a MTB is visible on the left; on the right side the graphene substrate is shown. \textbf{e}~d$I$/d$V$ and \textbf{f}~normalized amplitude A/T$_0$ of Frota fit as a function of sample bias and distance $x$ (nm) from the edge of the island. 
		STM/STS parameters: \textbf{b}~\SI[parse-numbers=false]{2.3 \times 4}{\nano\meter\squared}; \textbf{b,c}: $V_\text{set} = \SI{5}{\mV}$, $I_\text{set} = \SI{0.5}{\nA}$; \textbf{e}~\SI[parse-numbers=false]{5.7 \times 2.2}{\nano\meter\squared}, $V_\text{set} = \SI{100}{\mV}$, $I_\text{set} = \SI{5.0}{\pA}$; \textbf{e}~$V_\text{set} = \SI{10}{\mV}$, $I_\text{set} = \SI{1.0}{\nA}$. $V_\text{mod} = \SI{0.2}{\mV}$, $f_\text{mod} = \SI{907.0}{\Hz}$ for all spectra.
			\label{FigS3}}
\end{figure*}
The Kondo resonance decays rapidly when moving away from the mirror twin boundary. Moving across the boundary towards \mo, we see that the Kondo resonance is quickly suppressed. This is shown in Supplementary Figure~\ref{FigS3}a-c, where the amplitude of the resonance is expressed as the ratio of the Frota amplitude $A$ and the background conductance $T_0$, using the Frota function~\cite{Frota1992, Gruber2018}:
$$ \mathcal{F}(V) \propto T_0 + \Im\Bigg[A \times i\e^{i\phi}\sqrt{\frac{i\Gamma}{eV-E_\text{F}+i\Gamma}}\Bigg].
$$, where $\Im$ denotes the imaginary part of the function, $\phi$ is a form factor describing the asymmetry of the peak and the half width at half maximum of the line is given by $2.542\Gamma$. The resonance becomes undetectable around 1 nm away from the boundary. Similarly, we find that the Kondo resonance decays rapidly on graphene, see Supplementary Figure~\ref{FigS3}d-f.

In Supplementary Figure~\ref{SFigFrotaNRGL}~we show a comparison of the Kondo resonance as described by Frota and NRG. The good agreement confirms the Frota fit to be an adequate description of the Kondo resonance at the experimental temperatures.

	\begin{figure*}[h!]
		\centering
		\includegraphics[width=0.7\textwidth]{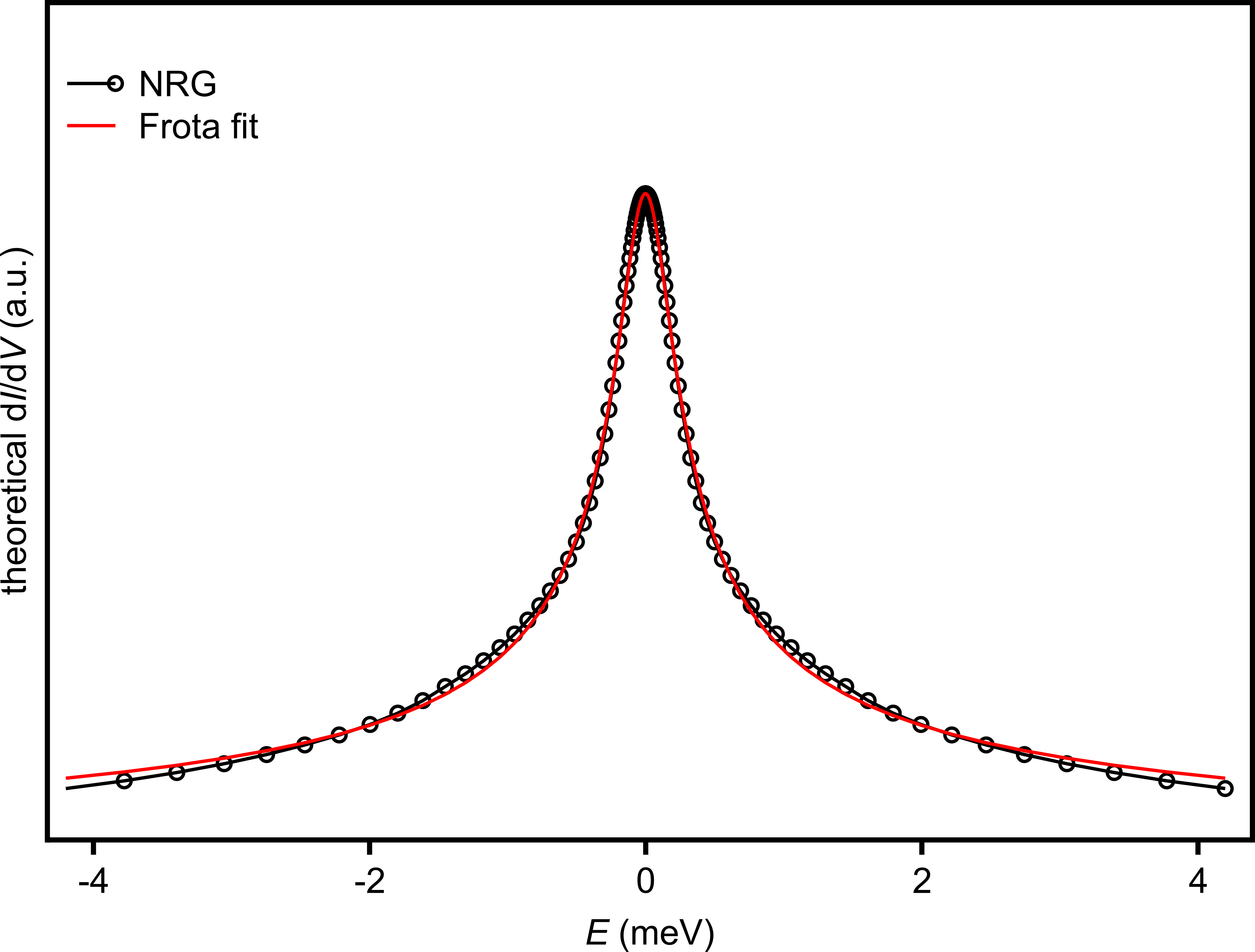}
		\caption{\footnotesize \textbf{Comparison between NRG and Frota.}~NRG was performed with $U = \SI{98}{\meV}$, $\gamma = \SI{9.35}{\meV}$, $T_\text{eff} =  \SI{0.7}{\K}$. The Frota fit to the NRG data used $T_0 = 0.00744$, $A = 0.00786$ and $\Gamma = \SI{0.1997}{\meV}$.
			\label{SFigFrotaNRGL}}
	\end{figure*}

\newpage

\section*{Supplementary Note 5: Anisotropy of the $g$-factor}
    \addcontentsline{toc}{section}{Supplementary Note 5: Anisotropy of the $g$-factor}

	\begin{figure*}[h!]
		\centering
		\includegraphics[width=0.5\textwidth]{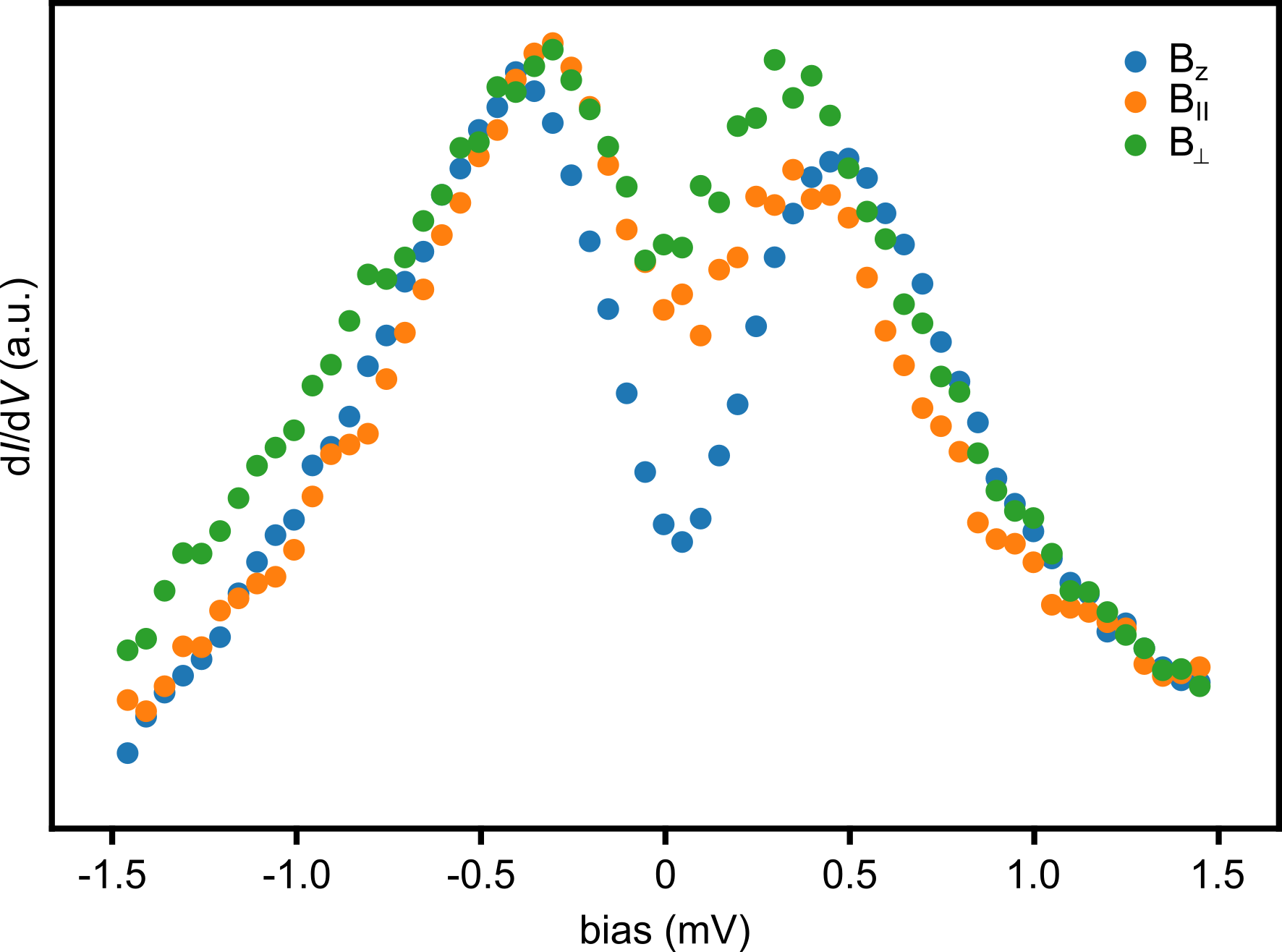}
		\caption{\footnotesize \textbf{Anisotropic splitting of the Kondo resonance under the influence of a magnetic field.}~Vectorial magnetic field-dependence of the Kondo resonance. $B_\text{z}$ points out-of-plane, while $B_{||}$ ($B_\perp$) points along (across) the mirror twin boundary. All spectra are taken at $|B| = \SI{0.75}{\tesla}$.
			\label{SFigAni}}
	\end{figure*}

Supplementary Figure~\ref{SFigAni} shows the splitting of the Kondo resonance for various field directions. The splitting is clearly largest for magnetic field pointing perpendicular to the \mo~plane ($B_z$), while we observe similar splitting for magnetic fields pointing in the \mo~plane parallel ($B_{||}$) and perpendicular ($B_{\perp}$) to the mirror twin boundary.

\newpage

\section*{Supplementary Note 6: Temperature dependence of Kondo resonance}
    \addcontentsline{toc}{section}{Supplementary Note 6: Temperature dependence of Kondo resonance}
	\begin{figure*}[h!]
		\centering
		\includegraphics[width=0.7\textwidth]{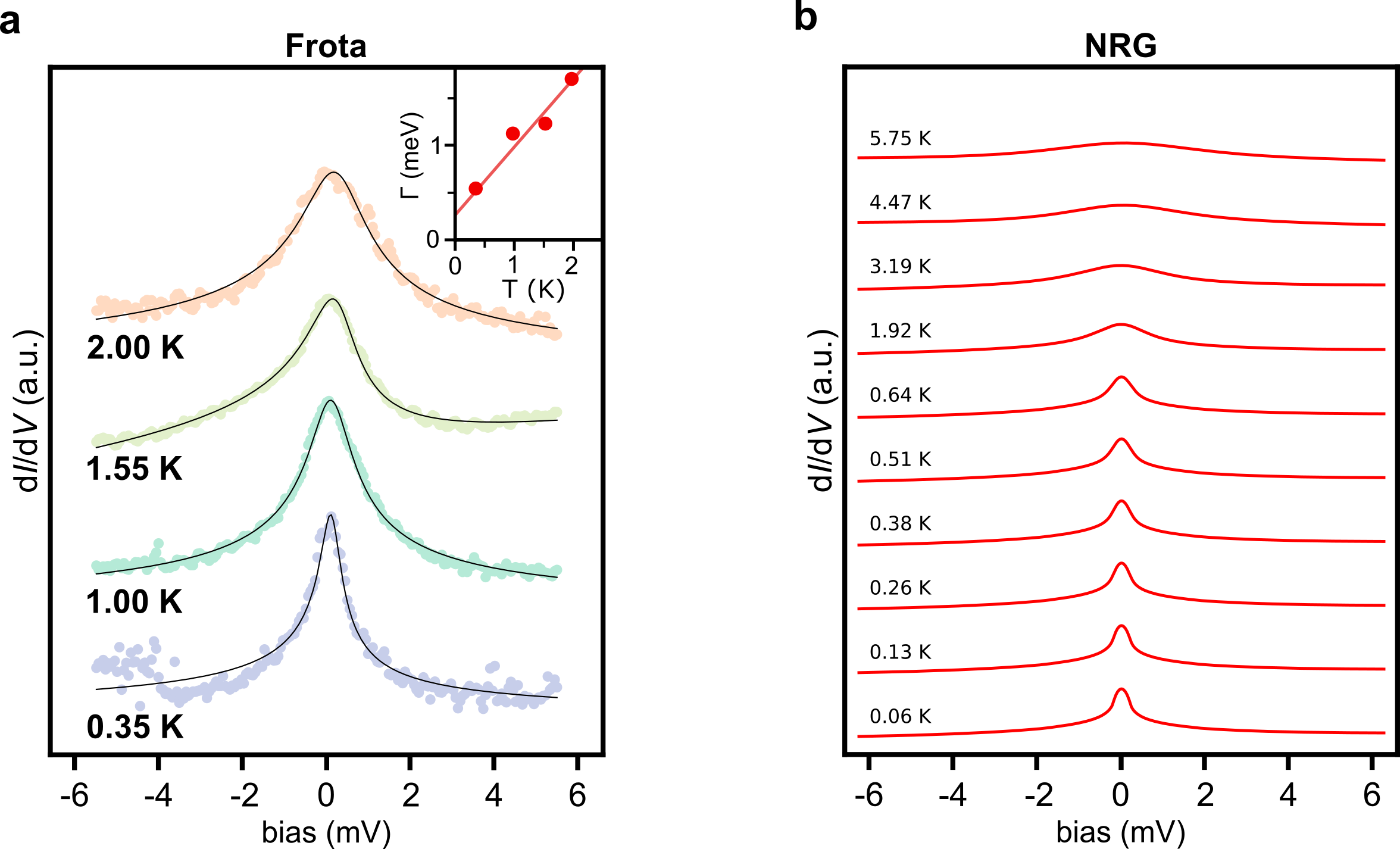}
		\caption{\footnotesize \textbf{Temperature dependence of Kondo resonance. a}~Dependence of the d$I$/d$V$ signal (averaged over the length a mirror twin boundary) on temperature ($B = \SI{0}{\tesla}$). The grey lines are a Frota function fitted to the experimental data, which is represented as dots. The inset displays in red dots the width $\Gamma (T)$ obtained from the Frota fit. The red line is a linear fit to $\Gamma (T)$. 
		\textbf{b}~Additional temperature-dependent NRG simulations for this specific boundary.
		STM/STS parameters:~$V_\text{set} = \SI{5}{\mV}$, $I_\text{set} = \SI{0.5}{\nA}$, $V_\text{mod} = \SI{0.2}{\mV}$
			\label{SFigTdep}}
	\end{figure*}
	
Supplementary Figure~\ref{SFigTdep}~presents additional fits to the temperature-dependence of the Kondo effect of the mirror twin boundary shown in Fig. 2 of the main text. Supplementary Figure~\ref{SFigTdep}a~represents a Frota fit (line) to the experimental data (dots). This Frota width is often used to extract the Kondo temperature $T_\text{K}$ by fitting $\Gamma (T)$ to $\Gamma = 2\sqrt{(\pi k_\text{B} T )^2 + 2(k_\text{B}T_\text{K})^2)}$~\cite{Nagaoka2002} (a discussion on Kondo scales is found in Supplementary Note 12). At temperatures well above $T_\text{K}$, this function decreases linearly with decreasing temperature, while at temperatures $T < T_\text{K}$ the width should saturate at a value $\propto T_\text{K}$~\cite{Cronenwett1998}. Since we find a linear temperature depedence down to our lowest temperature we can conclude that our lowest temperatures is still well above $T_\text{K}$, in line with our NRG calculations.

Supplementary Figure~\ref{SFigTdep}b shows the NRG prediction for a range of temperatures above and below the experimental values for this mirror twin boundary. Counter-intuitively, we find that the Kondo resonance is well visible up to $T\approx \SI{4}{\K}$, despite its small Kondo temperature of $T_\text{K} = \SI{2\times10^{-9}}{\K}$. NRG calculations at still higher temperatures show that for $T > \SI{45}{\K}$ the Kondo resonance is no longer discernible since the contribution at $E_\text{F}$ coming from the tails of the non-degenerate states overwhelms that from the Kondo effect.

\newpage

\section*{Supplementary Note 7: Anderson impurity model and parameter regimes}
    \addcontentsline{toc}{section}{Supplementary Note 7: Anderson impurity model and parameter regimes}
\label{sec:model}  

\begin{figure*}[h!]
  \centering 
  \includegraphics[width=0.8\linewidth]{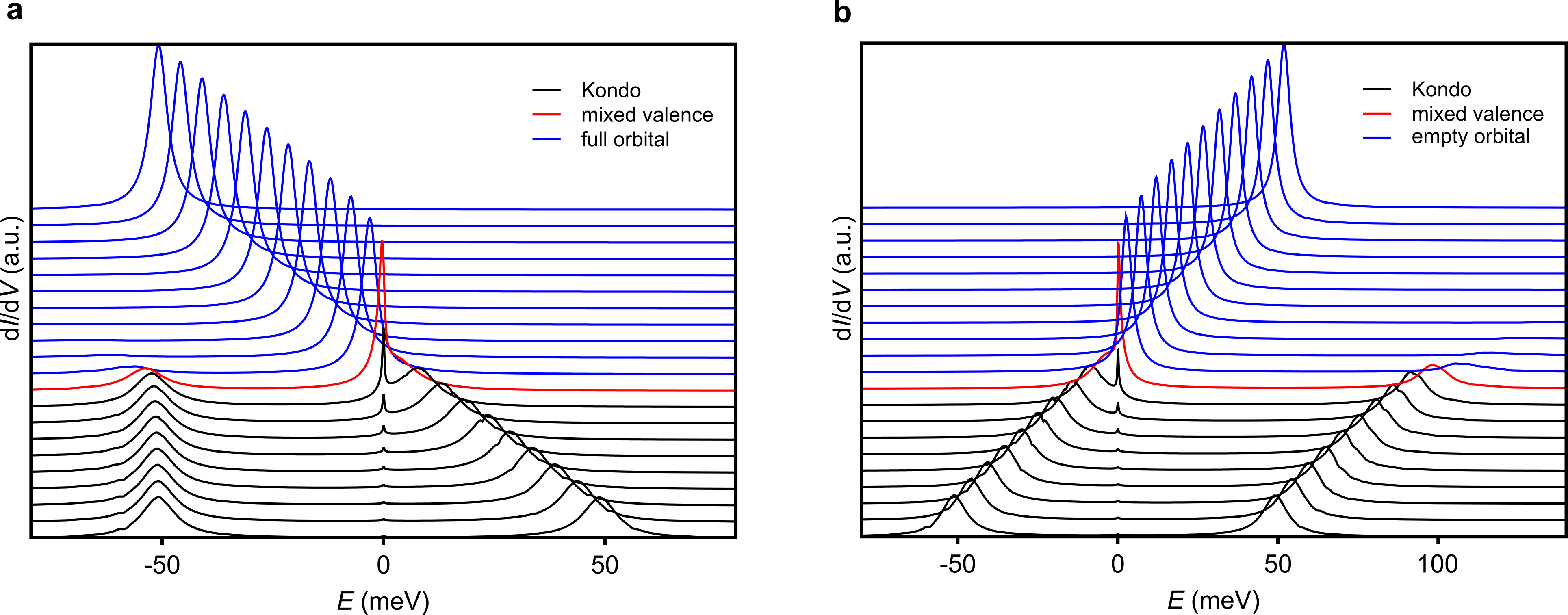}
\caption 
{\footnotesize \textbf{Evolution of d$I$d$V$ spectra with $U$ and $\varepsilon$ at $T=\SI{0.7}{\K}$ and $B = \SI{0}{\tesla}$ ($\gamma_0= \SI{4.68}{\meV})$. a}~Fixed $\varepsilon=\SI{-51}{\meV}$ and decreasing $U$ from $\SI{100}{\meV}$ (bottom) to $\SI{0}{\meV}$ (top), showing first a symmetric Kondo regime, then an asymmetric Kondo regime with a growing Kondo resonance, which eventually merges with the Hubbard peak at $\varepsilon+U$ to form a mixed valence resonance when $|\varepsilon+U|\lesssim \gamma_0/2$. Upon further reduction of $U$, a full orbital regime appears for $\varepsilon+U \le -\gamma_0/2$. 
  \textbf{b}~Fixed $U=\SI{100}{\meV}$ and increasing $\varepsilon$ from $\SI{-51}{\meV}\approx -U/2$ (bottom) to $\SI{+51}{\meV}\approx +U/2$ (top), showing first a symmetric Kondo regime, then an asymmetric Kondo regime with a growing Kondo resonance. The latter merges with the Hubbard peak at $\varepsilon$ to form a mixed valence resonance for $|\varepsilon|\lesssim \gamma_0/2$ (charge fluctuations between $ n=0$ and $n=1$). Empty orbital regime for $\varepsilon\ge \gamma_0/2$. NRG calculations: discretization parameter $\Lambda=5$, z-averaging with $N_z=192$ and retaining $1400$ states per iteration. A vertical offset of $0.1$ is used. 
}
\label{fig:fig-u+eps}
\end{figure*}

The NRG simulations of the d$I$/d$V$ spectra of the MTB's are based on the single-level Anderson impurity model, given by
\begin{align}
  H= & \sum_{\sigma}(\varepsilon -g\mu_BB\sigma/2) n_{\sigma} +Un_{\uparrow}n_{\downarrow} +\sum_{k\sigma}\varepsilon_{k}c_{k\sigma}^{\dagger}c_{k\sigma} + 
       \sum_{k\sigma}V_{k}(c_{k\sigma}^{\dagger}d_{\sigma} +d_{\sigma}^{\dagger}c_{k\sigma}).\label{eq:ham}
\end{align}
This describes a spin degenerate level of energy $\varepsilon$ relative to the Fermi energy, $E_\text{F}=0$, hybridizing with conduction electron states ($\varepsilon_{k}$) of the substrate via
hybridization matrix elements  $V_{k}$. In  (\ref{eq:ham}), $n_{\sigma}=d^{\dagger}_{\sigma}d_{\sigma}$ is the occupation number for spin $\sigma={\uparrow,\downarrow}$ electrons in the impurity level, $U$ is a local Coulomb repulsion and $B$ is a magnetic field acting on the impurity electrons via a Zeeman shift $-g\mu_BB\sigma/2$ with $g$ the g-factor. We take a constant $V_{k}=V$ and a constant conduction electron density of states $\rho(E)=\rho =1/2D$ with $-D\leq E \leq +D$, where $D$ is the half-bandwidth. Hence, the hybridization function $\gamma_0(E)=2\pi\sum_{k}|V_{k}|^2\delta(E-\varepsilon_{k})$ is a constant $\gamma_0(E)=\gamma_0=2\pi\rho V^2$ and equals the  full-width at half-maximum (FWHM) of the $U=0$ resonant level. The model is fully characterized by the parameters $U$, $\varepsilon$ and $\gamma_0$. 
For the MTB's of interest, the graphene monolayer on the iridium substrate is doped and has a metallic density of states at $E_\text{F}$ with the Dirac point lying far above \cite{Jolie2019a}. The Hubbard excitations at $\varepsilon$ and $\varepsilon+U$  also lie far from the Dirac point. Thus, to a first approximation, NRG calculations in the wide-band limit of the Anderson impurity model, \textit{i.e.}, $D\gg \text{max}(U, |\varepsilon|,|\varepsilon+U|,\gamma_0)$, are justified. We note also on our use of the term Hubbard excitations to describe the excitations at $\varepsilon$ and $\varepsilon+U$. These are otherwise known in the literature on the Anderson model as the satellite peaks in the spectral function \cite{Hewson1997}, or in the context of molecular junctions as the HOMO (highest occupied molecular orbital) and LUMO (lowest unoccupied molecular orbital) states, respectively. In the present experiment with MTBs, these are termed the non-degenerate states.

The model is a paradigm for describing strong correlations $U/\gamma_0\gg 1$. In this limit, a number of interesting regimes occur, such as the spin-fluctuation dominated Kondo regime for $-\varepsilon\gg \gamma_0/2$ and $(\varepsilon+U)\gg \gamma_0/2$, or the charge-fluctuation dominated mixed valence regime for $|\varepsilon|\lesssim \gamma_0/2$\cite{Hewson1997}. These exhibit  emergent low energy scales and a strong temperature and/or magnetic field dependence in physical quantities. In contrast, for weak correlations, $U/\gamma_0\lesssim 1$, renormalization effects are small, and the physics is essentially that of a noninteracting resonant level with a Lorentzian spectral function $A(E)=(\gamma_0/2\pi)/((E-\varepsilon)^2+(\gamma_0/2)^2)$. The MTB's in this paper are deep in the strongly correlated regime with $U/\gamma_0\sim 6-20$, see Fig.~3b in the main text. By comparison, semiconductor quantum dots exhibiting the Kondo effect typically have $U/\gamma_0\sim 3-6$ \cite{Goldhaber1998}.
Moreover, since the Hubbard peaks $-\varepsilon/\gamma_0\gg 1$ and $(\varepsilon+U)/\gamma_0 \gg 1$ are well separated from the Fermi energy, see Fig.~3e of the  main text, the impurity is singly occupied  and has $S=1/2$, i.e., the system  is in the Kondo regime. Other regimes can, in principle, be realized by pulsing or gating the MTB to shift $\varepsilon$ relative to
the Fermi energy. For example, when $\varepsilon$ (or $\varepsilon+U$) is within $\gamma_0/2$ of $E_\text{F}=0$ the mixed valence regime is realized. When
$\varepsilon \gg \gamma_0/2$ (or $\varepsilon+U \ll -\gamma_0/2$) the  empty (or full) orbital regime emerges, which despite having a large $U/\gamma_0$ is effectively noninteracting because the empty (or full orbital) cannot undergo charge or spin fluctuations. However, as discussed in more detail at the end of this note, for such large $\varepsilon$ additional levels in the MTB's need to be considered, so the simple empty (full) orbital regime is not realizable.

Figure~\ref{fig:fig-u+eps} illustrates the parameter regimes of the model in terms of the trends in d$I$/d$V$, calculated within NRG, upon decreasing $U$ for a fixed $\varepsilon$ (Figure~\ref{fig:fig-u+eps}a) or upon increasing $\varepsilon$ for a fixed $U$ (Figure~\ref{fig:fig-u+eps}b). We take as reference system the MTB in Fig.~2 of the main text with the parameters given above (but allowing either $U$ or $\varepsilon$ to vary) and the experimental temperature $T=0.7 \text{ K}$ and zero magnetic field. Decreasing $U$ at fixed $\varepsilon\ll E_\text{F}$ drives the system from the Kondo regime (black lines) through a mixed valence regime for $|\varepsilon+U|\lesssim \gamma_0/2$ (red line), a full orbital regime for $\varepsilon+U \le \gamma_0/2$ with $U$ finite and eventually to the noninteracting resonant level regime at $U=0$ (Figure~\ref{fig:fig-u+eps}a). Note how the full orbital resonance with finite $U$ continuously connects to the noninteracting ($U=0$) resonant level.  Similarly, with $U$ fixed and increasing $\varepsilon$, the system is driven from the Kondo and asymmetric Kondo regimes to a mixed valence regime for $|\varepsilon|\lesssim\gamma_0/2$ and eventually to the empty orbital regime for $\varepsilon >\gamma_0/2$ (Figure~\ref{fig:fig-u+eps}b).
Notice how the Kondo resonance grows in the asymmetric Kondo regime as $\varepsilon$ (or $\varepsilon+U$) approaches $E_\text{F}$, eventually merging with this level to form the mixed valence resonance when
$|\varepsilon|\lesssim \gamma_0/2$ (or $|\varepsilon+U|\lesssim \gamma_0/2$).  Supplementary Figure~2d (Supplementary Note 2) shows a Hubbard excitation moving closer to the Fermi energy under pulsing. This could eventually give rise to a mixed valence resonance at $E_\text{F}$, either when $\varepsilon$ or $\varepsilon+U$ is within $\gamma_0/2$ of $E_\text{F}$.

The trends in the {\em weights} of the high energy Hubbard excitations in d$I$/d$V$ can be easily understood within the exactly solvable atomic limit $V=0$. This gives a two peaked  spectral function  $A(E,T) = (2-n)\delta(E-\varepsilon) + n\delta(E-(\varepsilon+U))$, where $n$ is the impurity occupation.  The latter changes from
$n=1$ (symmetric Kondo regime $\varepsilon=-U/2$) to $n=2$ (full orbital regime $\varepsilon+U\le -\gamma_0/2$) on decreasing $U$ at fixed $\varepsilon$ (Fig.~\ref{fig:fig-u+eps}a) and from $n=1$ (symmetric Kondo regime) to $n=0$ (empty orbital regime) on increasing $\varepsilon$ at fixed $U$ (Fig.~\ref{fig:fig-u+eps}b). 

A finite $V>0$ results, (i), in a finite width for the Hubbard peaks $\propto V^2$ and, (ii), in a low energy Kondo resonance appearing at $E_\text{F}$, generated by $O(V^4)$ spin-flip processes. As discussed in the main text and in Supplementary Note~8, in the Kondo regime (black curves) the full width $\gamma$ of the Hubbard peaks at $\varepsilon$ and $\varepsilon+U$ is {\em twice} the bare value $\gamma_0$ as a result of spin-flip processes \cite{Pruschke1989,Logan1998}, i.e.,
$\gamma=2\gamma_0$. This is important for interpreting the experiments which measure the enhanced width $\gamma$, whereas the input parameter for the model calculations is $\gamma_0=\gamma/2$. This factor of two enhancement of the width is absent in the empty (or full) orbital regime which is effectively noninteracting so in this case $\gamma=\gamma_0$ (Figs.~\ref{fig:fig-u+eps}a,b). The width of the mixed valence resonance is also of order $\gamma_0$,  but its position is strongly renormalized \cite{KWW1980b}.

In real MTB's additional levels enter which will modify the above picture. The situation we described with the single level Anderson model assumes that all other levels are fully
  occupied or empty and do not enter the transport window as $\varepsilon$ is changed. This assumption remains valid in the Kondo regime and in the mixed valence regime, 
  but will break down in the empty (full) orbital regime when additional levels start to be populated (de-populated) and thus contribute to transport.

\newpage

\section*{Supplementary Note 8: Broadening of the impurity states in the spectral function}
    \addcontentsline{toc}{section}{Supplementary Note 8: Broadening of the impurity states in the spectral function}

\begin{figure*}[h!]
		\centering
		\includegraphics[width=0.7\textwidth]{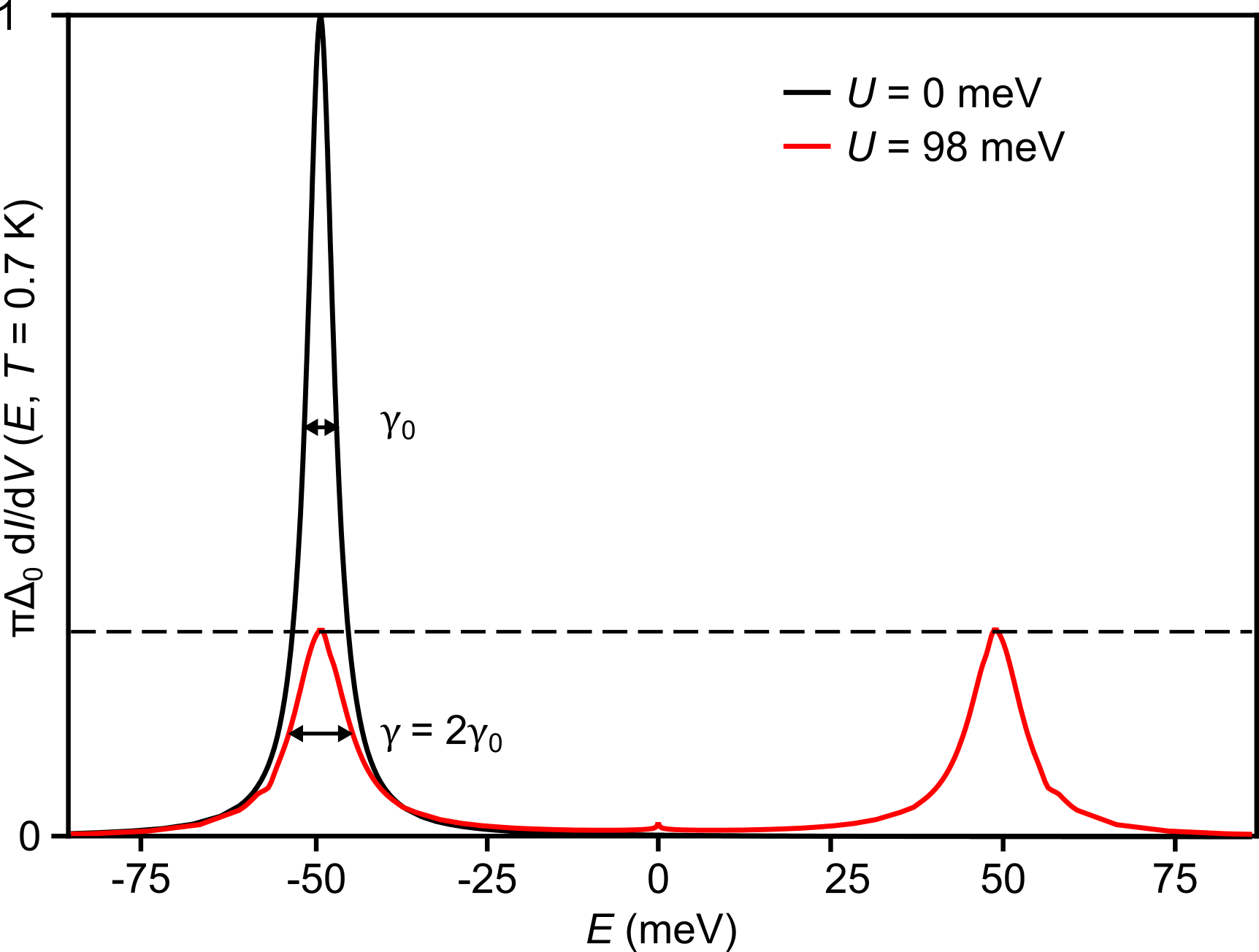}
		\caption{\footnotesize \textbf{Broadening of the impurity states.}~Influence of the impurity peak position on the broadening and Kondo resonance. NRG calculations performed with $T_\text{eff} = \SI{0.7}{\K}$, $\varepsilon = \SI{-49}{\meV}$, $\gamma =  \SI{9.35}{\meV}$.
			\label{SFigbroad}}
	\end{figure*}
	
The impurity states have a finite width in the spectral function. This width is composed of two terms: (i) broadening caused by hybridization of impurity states with the electron bath of the substrate, and (ii) spin-flip processes of the electron occupying the impurity state~\cite{Logan1998}. While both are always present in an experimental Kondo system, NRG is able to separate the two by setting $U=0$, shown in Supplementary Figure~\ref{SFigbroad}. In this limit, only hybridization contributes to the broadening of the impurity states, leading to a FWHM of $\gamma_0$. In comparison, we find that a sizeable $U$~redistributes the spectral weight on the lower and uppper Hubbard impurity levels, leading to a reduction of the peak height by a factor of 2. An additional reduction is due to to spin-flip processes, leading to a peak height reduction by a factor of 4 and a peak width of $\gamma=2\gamma_0$.

\newpage

\section*{Supplementary Note 9: Comparison of spectral function and d$I$/d$V$~for a mirror twin boundary with asymmetric states}
    \addcontentsline{toc}{section}{Supplementary Note 9: Comparison of spectral function and d$I$/d$V$~for a mirror twin boundary with asymmetric states}

	\begin{figure*}[h!]
		\centering
		\includegraphics[width=\textwidth]{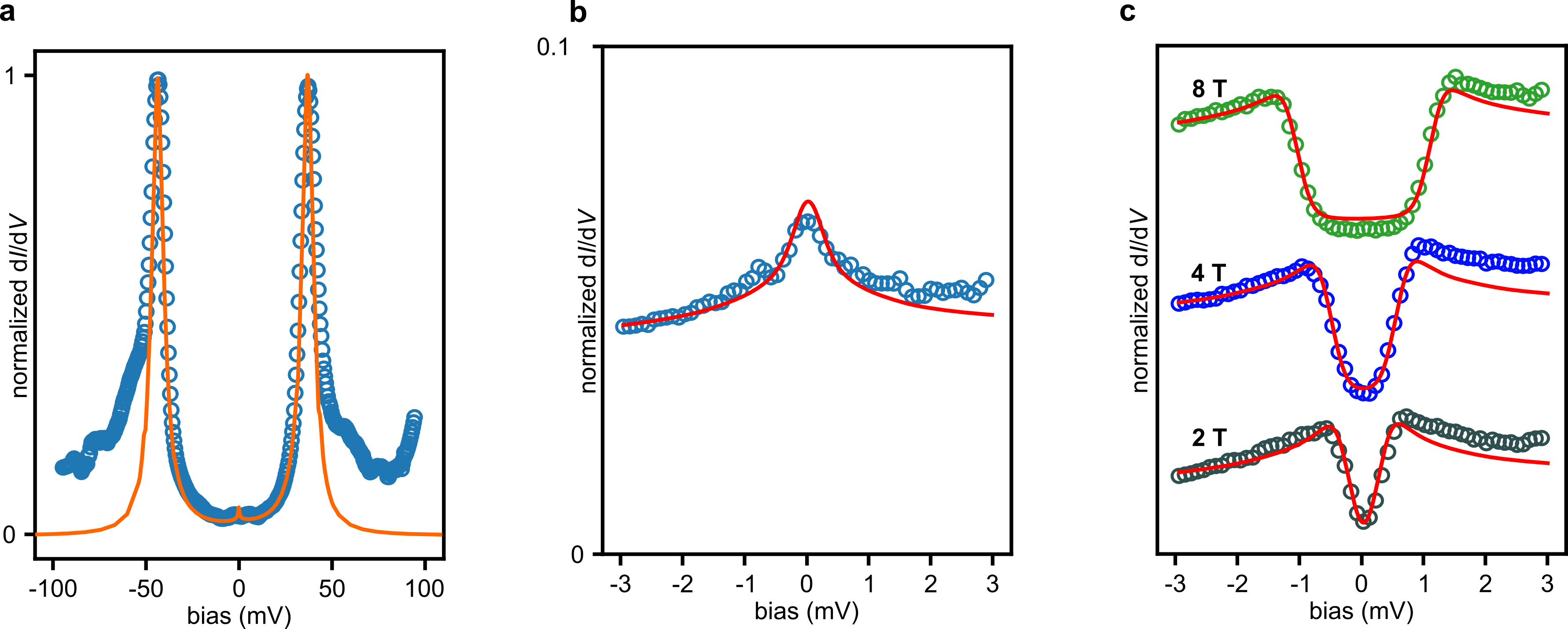}
		\caption{\footnotesize \textbf{Comparison between NRG and d$I$/d$V$ for a mirror twin boundary with asymmetric states. a}~d$I$/d$V$ spectrum measured on a MTB with $L = \SI{11.2}{\nm}$, $\varepsilon = \SI{-44}{\meV}$, $U = \SI{81}{\meV}$, $\gamma_\text{NRG} = \gamma = \SI{7.8}{\meV}$, with a NRG fit to the data (orange line). Data and fit normalized to highest peak. \textbf{b}~Kondo resonance measured with STS (blue circles) with NRG fit (red line), plotted in the same scale as \textbf{a}. Note that NRG is not fitted to the Kondo resonance, only to the non-degenerate states. \textbf{c}~Dependence of d$I$/d$V$ signal on magnetic field ($T_\text{eff} = \SI{0.7}{\K}$), with NRG data for $g = 2.5$.
			\label{SFigasym}}
	\end{figure*}

Supplementary Figure~\ref{SFigasym} presents a comparison between theoretical NRG simulations and experimental d$I$/d$V$ spectra for a mirror twin boundary with asymmetric states around the Fermi energy, i.e., $\varepsilon\neq -U/2$. Due to the reduced inelastic tails, we find that for $\gamma_\text{NRG} =  \gamma$, NRG correctly predicts the intensity, width and magnetic field dependence of the Kondo resonance, using $g=2.5$ in the simulations.

\newpage

\section*{Supplementary Note 10: Dependence of quantization energy and Coulomb gap on boundary length}
    \addcontentsline{toc}{section}{Supplementary Note 10: Dependence of quantization energy and Coulomb gap on boundary length}

	\begin{figure*}[h!]
		\centering
		\includegraphics[width=\textwidth]{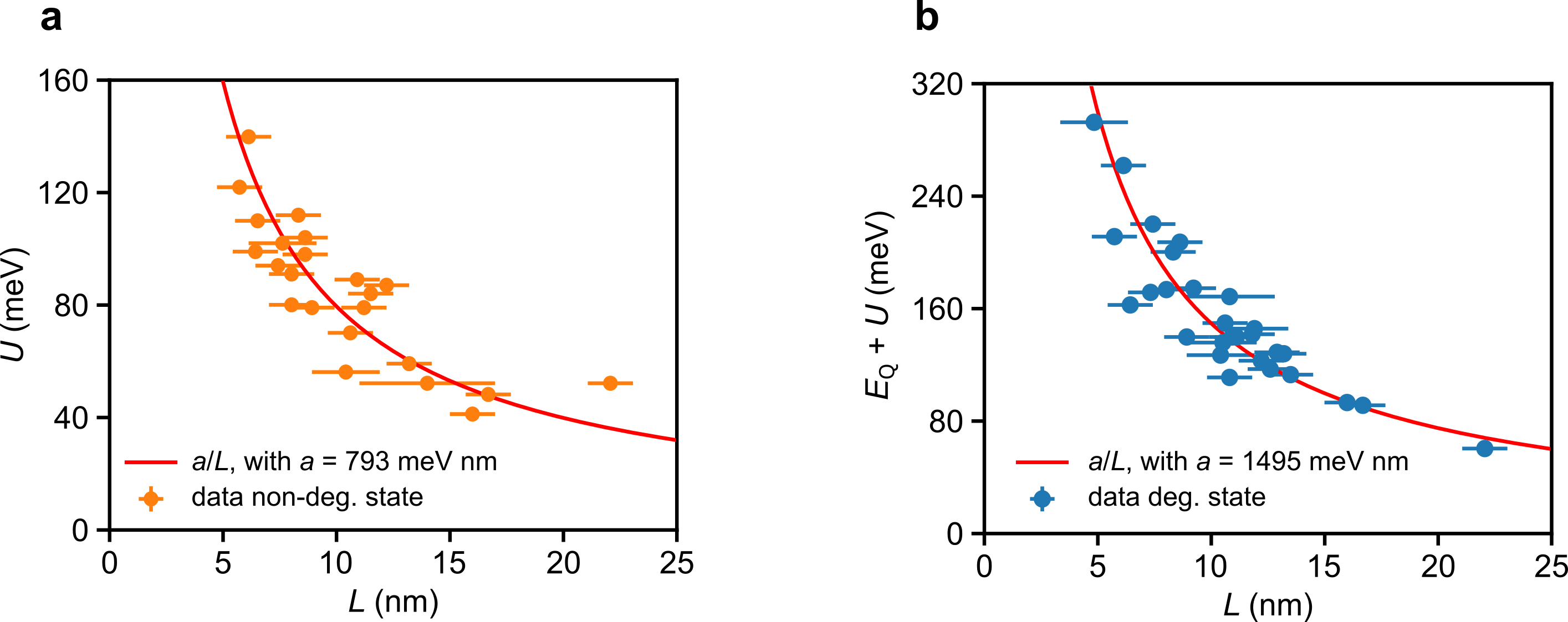}
		\caption{\footnotesize \textbf{Coulomb gap ($U$) and degenerate gap $E_\text{gap} = E_\text{Q} + U$ dependence on boundary length ($L$). a}~$U$ versus $L$. \textbf{b}~\textbf{b}~$E_\text{Q} + U$ versus $L$. Fits $y = a/L$ are shown in red.
			\label{SFigUvsL}}
	\end{figure*}

The quantization energy $E_\text{Q}$ of a one-dimensional system with linear dispersion is $\hbar vq_n$, with $v$ the velocity and $q_n = (\pi/L)n$~\cite{Jolie2019a}. Likewise, also the Coulomb energy $U$ is proportional to $1/L$~\cite{Zhu2021}. As observed previously in Refs.~\citenum{Jolie2019a, Zhu2021}, we indeed find a $1/L$ dependence of $U$, see Supplementary Figure~\ref{SFigUvsL}a. The sum of the quantization and the Coulomb energy, which makes up the gap $E_\text{gap} = E_\text{Q} + U$ that we find in the non-degenerate state, also scales with $1/L$, as plotted in Supplementary Figure~\ref{SFigUvsL}b. 

\newpage

\section*{Supplementary Note 11: Dependence of $\gamma$ on fitting procedure}
    \addcontentsline{toc}{section}{Supplementary Note 11: Dependence of $\gamma$ on fitting procedure}
\begin{figure*}[h!]
		\centering
		\includegraphics[width=0.9\textwidth]{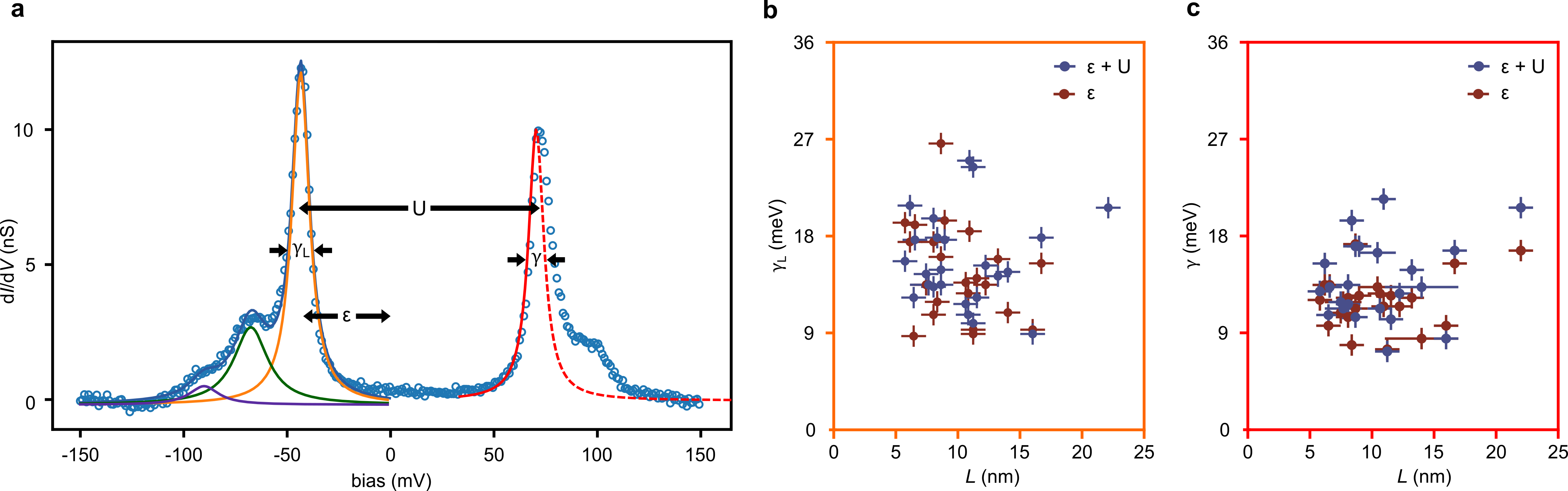}
		\caption{\footnotesize \textbf{Dependence of hybridization $\gamma$ on fitting procedure. a}~d$I$/d$V$ spectrum of non-degenerate states (blue circles) of a MTB with $L = \SI{8.6}{\nm}$, $\varepsilon = \SI{-51}{\meV}$, $U = \SI{100}{\meV}$. On the left state, a Lorentzian fit (dark blue line) with three components is shown: the non-degenerate peak (orange line) and two phonon satellites (green and purple lines). Indicated are the Coulomb energy $U$, the energy spacing $\varepsilon$~from $E_\text{F}$, and the full width at half maximum of the main Lorentzian peak $\gamma_L$. On the right state, a single Lorentzian fitted to the inner slope of the peak is shown, giving $\gamma$. \textbf{b}~FWHM of non-degenerate states obtained by a fit with three skewed Lorentzian functions $\gamma_\text{L}$. \textbf{c}~FWHM $\gamma$ of Lorentzians fitted to inner slope of non-degenerate states $\gamma$. STM/STS parameters: \textbf{a}~$V_\text{set} = \SI{200}{\mV}$, $I_\text{set} = \SI{0.2}{\nA}$, $V_\text{mod} = \SI{1.0}{\mV}$.
			\label{SFigGVar}}
	\end{figure*}

The phonon excitations caused by inelastic tunneling processes lead to an artificial broadening of the confined peaks in experiment. This effect is strongest at absolute energies larger than the peak energies, leading to the appearance of inelastic tails towards higher absolute energies. Supplementary Figure~\ref{SFigGVar}a compares two ways to extract the width of the confined states from the experimental d$I$/d$V$ spectra. One is to fit the non-degenerate states with up to three Lorentzians, dependent on the number of visible phonon peaks. This gives us a width $\gamma_L$ of which the distribution of values is shown in Supplementary Figure~\ref{SFigGVar}b. The method applied in the main manuscript  (Fig. 3e), on the other hand, is to fit the inner tail of the peaks using a single Lorentzian with width $\gamma$, shown in Supplementary Figure~\ref{SFigGVar}c. Both distributions do not show a correlation with the mirror twin boundary length $L$, but $\gamma_L$ tends to be larger than $\gamma$ for the same peak. This is because the inelastic tail is not only due to discrete phonon losses at the energies of van Hove singularities in the phonon dispersion, but by phonon losses with the entire spectrum of phonon energies in the phonon dispersion. Therefore we use $\gamma$ obtained from fitting to the inner slope as a more reliable estimate based on our direct comparison with NRG, instead of $\gamma_L$. Note that for both $\gamma$ and $\gamma_L$, we fit both the upper and lower peak and use the average peak width $\gamma = (\gamma_\text{lower} + \gamma_\text{upper})/2$ in the Anderson model. 
	
\newpage	

\section*{Supplementary Note 12: Kondo scales}
    \addcontentsline{toc}{section}{Supplementary Note 12: Kondo scales}
\label{sec:Kondo-scalesl}
The two commonly used definitions for the Kondo scale in the Anderson and Kondo models, are the low temperature strong coupling Kondo scale, denoted by $k_\text{B}T_0$, and the perturbative Kondo scale $k_\text{B}T_\text{K}$, related to the former via $T_\text{K}/T_0=w$, where the universal Wilson number $w = e^{C+1/4}/\pi^{3/2}=0.4128$ with $C=0.577216$ being Euler's constant \cite{Hewson1997}. In limiting cases,
analytic expressions for $T_\text{K}$ and $T_0$ can be given for the Kondo and Anderson models, as described below.

Besides these two scales, a third scale is of particular interest in STM, namely the HWHM $\Gamma_\text{K}$ of the zero temperature Kondo resonance in the spectral function or in d$I$/d$V$, 
\begin{align}
\frac{\text{d}I}{\text{d}V}(\SI{}{\eV}) & = \int_{-D}^{+D}\frac{A(E,T)}{4k_\text{B}T \cosh^2((E-eV)/2k_\text{B}T)}.\label{eq:NRG-dIdV}
\end{align} 
More precisely, denoting by $\Gamma^A(T)$ and $\Gamma^G(T)$ the HWHM of the Kondo peak in $A(E,T)$ and  $G(V,T)=\text{d}I/\text{d}V$ at temperature $T$, respectively, one clearly has
in the zero temperature limit $\Gamma^G(T=0)=\Gamma^A(T=0)=\Gamma_\text{K}$. At finite temperature, $\Gamma^A(T)$ and  $\Gamma^G(T)$  will differ due to the additional thermal broadening in
d$I$/d$V$, i.e., $\Gamma^G(T)\ge \Gamma^A(T)$. Furthermore, the experimentally measured differential conductance $G_{\text{expt}}(eV,T)$ will be a convolution of the intrinsic d$I$/d$V$ and the
resolution in the measurement protocol (described in Methods in the main text). Deconvoluting allows to extract $\Gamma^A(T)$ and  $\Gamma^G(T)$ from the  measured $G_{\text{expt}}(eV,T)$ \cite{Gruber2018}. Theoretically, however, both $\Gamma^G(T)$ and $\Gamma^A(T)$ can be accessed directly. Below, we  calculate the universal curves for $\Gamma^G(T)$ and $\Gamma^A(T)$ as a function of $T/T_0$ and show that
$\Gamma_\text{K} = \pi k_\text{B}T_0/2$ deep in the symmetric Kondo regime of the Anderson model for $U/\gamma_0\gg 1$. Hence, the relation between the three scales in this limit is
\begin{align}
  \Gamma_\text{K} & = \frac{\pi k_\text{B}T_0}{2} = \frac{\pi k_\text{B} T_\text{K}}{2w}. \label{eq:Kondo-scales-relation}
\end{align}

\subsection{Kondo model scales}
\label{subsec:km-scales}

\begin{figure*}[h!]
  \centering 
  \includegraphics[width=0.95\linewidth]{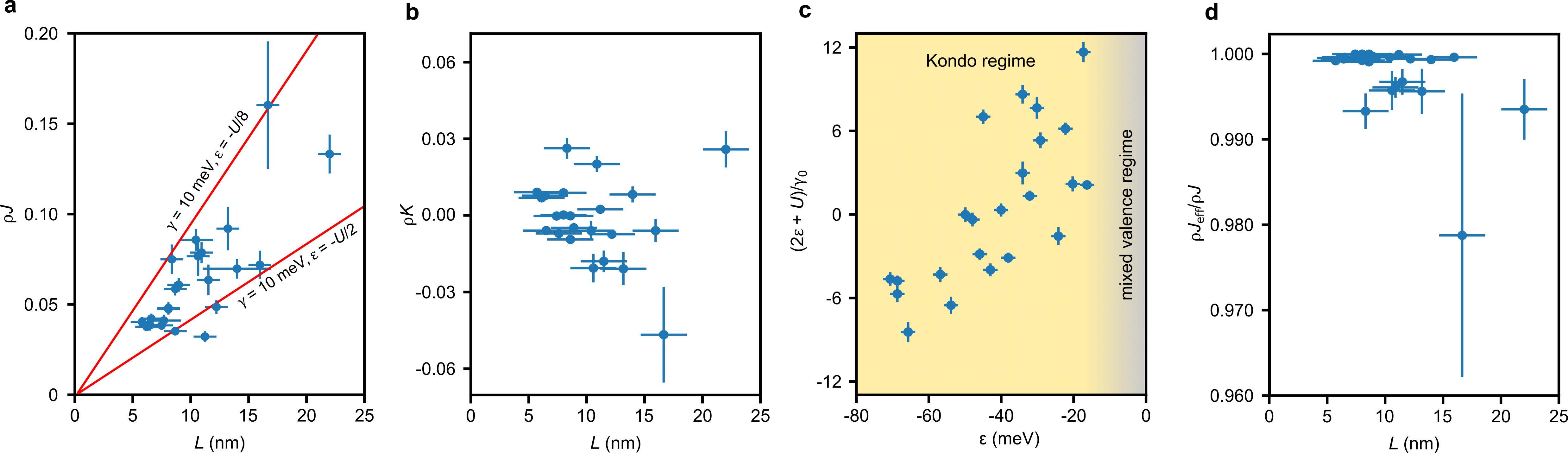}
\caption 
{\footnotesize \textbf{Kondo coupling  and potential scattering strengths. a~}Kondo coupling $\rho J$ from Eq.~(\ref{eq:SW1}) versus $L$ for the boundaries in the main text. The lines simulate $\rho J$ for the symmetric
  case $\varepsilon(L)=-U(L)/2$ and for an asymmetric case $\varepsilon(L)=-U(L)/8$ using constant $\gamma=10 \text{meV}$ and the $U(L)$ from Fig.3 of the main text. As expected, $\rho J$ increases away from particle-hole symmetry. {\bf b} Potential scattering strength $\rho K$ from Eq.~(\ref{eq:SW2})  versus $L$.
 {\bf c} Measure of the particle-hole asymmetry $\delta(\varepsilon)=(2\varepsilon+U)/\gamma_0$ versus $L$. All boundaries are in the Kondo regime (shaded yellow). The mixed valence regime $|\varepsilon|\lesssim \gamma_0/2$ is shaded grey. {\bf d} Ratio $\rho J_{\text{eff}}/\rho J$ versus $L$ showing that potential scattering has a negligible effect on the Kondo coupling. 
}
\label{fig:Kondo}
\end{figure*}

The low energy behaviour of the Anderson model in the  Kondo regime, i.e., for  $-\varepsilon/\gamma_0\gg 1$ and $(\varepsilon+U)/\gamma_0 \gg 1$, can be captured by the Kondo model $H_{K}= \sum_{k\sigma}\varepsilon_{k}c_{k\sigma}^{\dagger}c_{k\sigma} +2J\vec{S}\cdot\vec{s_{0}} + K\sum_{\sigma,\sigma'}c_{0\sigma}^{\dagger}c_{0\sigma'}$ with an effective  antiferromagnetic exchange coupling $J$ and a potential scattering term $K$ related to the Anderson model parameters ($U,\varepsilon$ and $\gamma_0$) by the Schrieffer-Wolff expression \cite{Schrieffer1966,Hewson1997} 
\begin{align}
  \rho J = & \frac{\gamma_0}{2\pi}\left(\frac{1}{\varepsilon+U} - \frac{1}{\varepsilon} \right) =\frac{\gamma_0U}{2\pi||\varepsilon|\varepsilon+U|},\label{eq:SW1}\\
\rho K = & -\frac{\gamma_0}{4\pi}\left(\frac{1}{\varepsilon+U}+\frac{1}{\varepsilon}\right) = \frac{\gamma_0^2}{4\pi}\frac{\delta(\varepsilon)}{|\varepsilon||\varepsilon+U|},\label{eq:SW2}
\end{align}
where $\delta(\varepsilon)=(2\varepsilon+U)/\gamma_0$ measures the particle-hole asymmetry in the two models. The latter is zero at the symmetric point $\varepsilon=-U/2$,
but is finite otherwise. The values of $\rho J$ and $\rho K$ for the boundaries in the main text are shown in Figs.~\ref{fig:Kondo}a,b together with 
$\delta(\varepsilon)$, a measure of their particle-hole asymmetry (Fig.~\ref{fig:Kondo}c). The latter show that the boundaries in the main text realize both the nearly particle-hole symmetric
Kondo regime ($\varepsilon\approx -U/2$) where $\delta(\varepsilon)\approx 0$ and also the asymmetric Kondo regime where $|\delta(\varepsilon)|>0$.

A finite $K$ has two effects, (i), it shifts the position of the Kondo resonance to either slightly above or slightly below the Fermi energy, depending on the sign of $\delta(\varepsilon)$,
and, (ii), it affects the Kondo scale, defined below, by renormalizing $\rho J\to \rho J_{\text{eff}}=\rho J/(1+(\pi\rho K)^2)$ \cite{KWW1980b}. The latter effect, however, is very weak, since
the potential scattering strength enters only quadratically in $\rho J_{\text{eff}}$. For the systems considered in the main text, all in the Kondo regime, the effect of potential scattering on the
Kondo coupling,  shown explicitly in Fig.~\ref{fig:Kondo}d, is negligible and we have $\rho J_{\text{eff}}\approx \rho J $. Thus, even in the presence of a finite potential scattering,
the usual expression for $T_\text{K}$ of the Kondo model applies,
\begin{align}
  k_\text{B}T_{K} & = D \sqrt{2\rho J}e^{-1/2\rho J + O (\rho J) }\approx D \sqrt{2\rho J}e^{-1/2\rho J  }.\;\;\;\text{for $\rho J\ll 1$}. \label{eq:TK-KondoModel-HighTdef}
\end{align}
The strong coupling scale can be obtained from this as $k_\text{B}T_0=k_\text{B}T_\text{K}/w$.

Note that from (Eq.~\ref{eq:SW1}), the measure for strong correlations in the Anderson model $U/\pi\gamma_0/2\sim (4/\pi) U/\gamma \sim U/\gamma \gg 1$ \cite{Hewson1997} translates, in the symmetric case $\varepsilon=-U/2$, to $U/\gamma \sim \frac{U}{\pi\gamma_0/2}=\frac{8}{\pi^2}\frac{1}{2\rho J} \approx \frac{1}{2\rho J}\gg 1$, i.e., strong correlations in the Kondo model correspond to $2\rho J \ll 1$.  This condition holds for the boundaries in the main text, both in terms of $U/\gamma_0$ (see main text Fig.3{\bf d}) and in terms
of the Kondo coupling $\rho J$ in Fig.~\ref{fig:Kondo}a which shows that $0.04 \lesssim \rho J \lesssim 0.16$.

Finally, note that the factor of 2 in the dimensionless Kondo coupling $2 \rho J$ holds for the above convention for the Kondo interaction ($2J\vec{S}\cdot\vec{s_{0}}$). This is used, for example, in the  work of Kondo \cite{Kondo1964} and in Ref.~\cite{Hewson1997}. Wilson\cite{Wilson1975} used the same notation with the opposite sign for $J$. The notation without the factor of $2$ is also common in the literature \cite{KWW1980a,KWW1980b}.

\subsection{Anderson model scales}
\label{subsec:am-scales}
\begin{figure*}[h!]
  \centering 
  \includegraphics[width=0.35\linewidth]{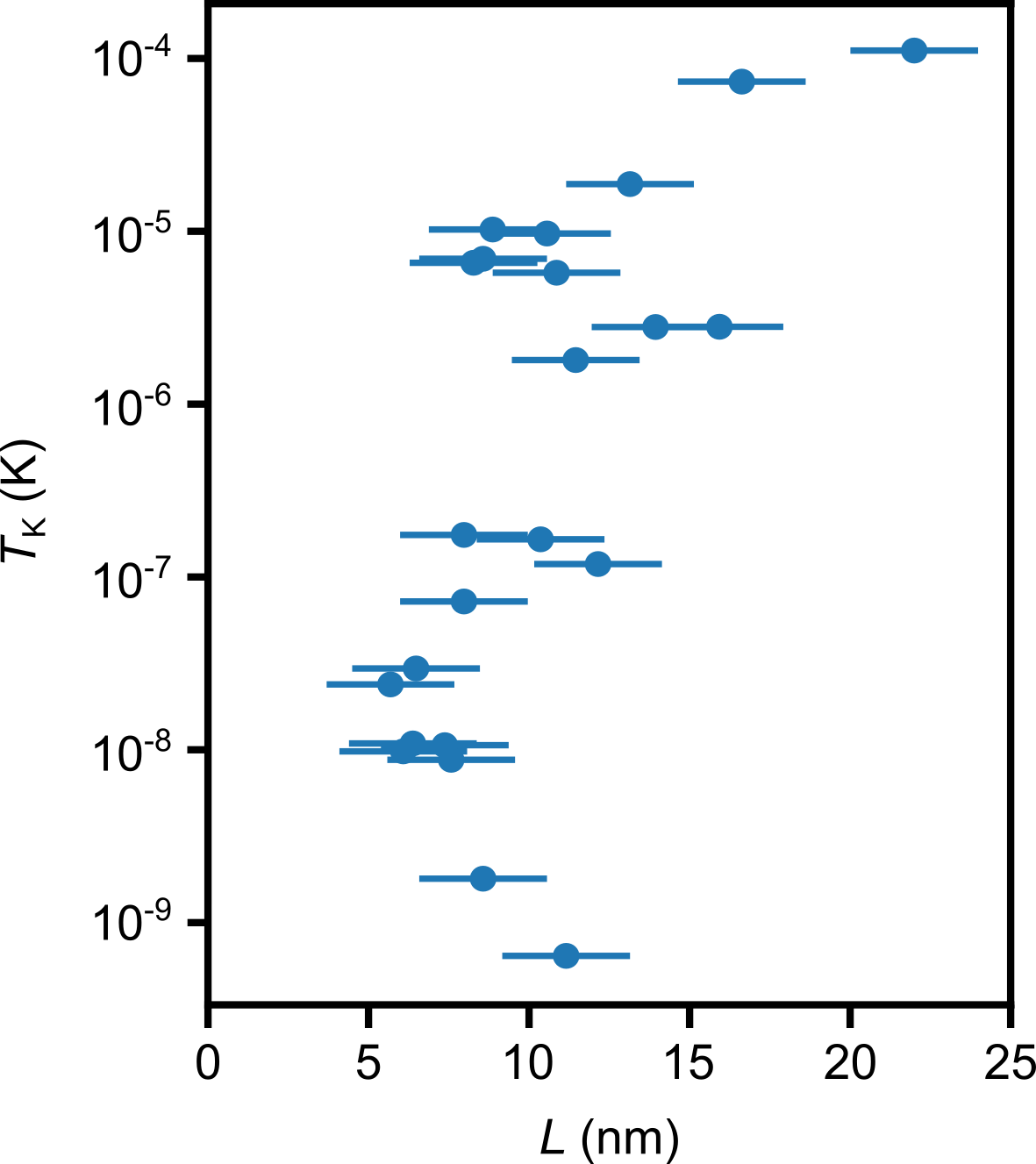}
\caption 
{\footnotesize \textbf{Kondo temperature $k_\text{B}T_\text{K}= w T_0=w \gamma_0 \sqrt{U/4\gamma_0}e^{-\pi|\varepsilon||\varepsilon+U|/U\gamma_0}$ for the boundaries in the main text.} The parameters $U,\varepsilon,\gamma$ are extracted from the d$I$/d$V$ spectra.  In terms of $\rho J$, this $T_\text{K}$ is equivalent to  Eq.~(\ref{eq:TK-AndersonModel-HighTdef}).
}
\label{fig:KondoScaleAM}
\end{figure*}

For the Anderson model,  $k_\text{B}T_0$ can be calculated accurately within NRG via its definition $\chi(T=0) =(g\mu_B)^2/4k_\text{B}T_0$. 
Analytic formulae are also available in limiting cases for $U/\gamma_0\gg 1$. For example, for $\varepsilon=-U/2$  Bethe-Ansatz gives \cite{Hewson1997}
\begin{align}
  k_\text{B}T_0 & = \gamma_0 \sqrt{U/4\gamma_0}e^{-\pi U/4\gamma_0 + \pi\gamma_0/4U},\;\;\;\text{for $\varepsilon=-U/2 \text{ and }U/\gamma_0\gg 1$},\label{eq:t0-symm}
\end{align}
whereas for the asymmetric Kondo regime, $\varepsilon\neq -U/2$, we have, to leading order in the exponent, 
\begin{align}
  k_\text{B}T_0 & = \gamma_0 \sqrt{U/4\gamma_0}e^{-\pi|\varepsilon||\varepsilon+U|/U\gamma_0}.\;\;\;\text{for $\varepsilon/\gamma_0 \ll -1, (\varepsilon+U)/\gamma_0\gg +1$ and  $U/\gamma_0\gg 1$}.\label{eq:t0-asymm}
\end{align}
All parameters appearing in (\ref{eq:t0-symm}) and (\ref{eq:t0-asymm}) can be measured  
in the d$I$/d$V$ of the present experiment, thereby allowing $k_\text{B}T_0$  and hence also $k_\text{B}T_\text{K}=wk_\text{B}T_0$ to be calculated.
Figure~\ref{fig:KondoScaleAM} shows $T_{K}$ calculated in this way for
the boundaries in the main text using the experimentally extracted parameters $U,\varepsilon,\gamma=2\gamma_0$.

While the Anderson model parameters suffice to determine $k_\text{B}T_\text{K}$ directly, one can also express the latter in terms of the Kondo coupling $\rho J$ analogously to the expression for $T_\text{K}$ in terms of $\rho J$ for the Kondo model in Eq.(\ref{eq:TK-KondoModel-HighTdef}). In the symmetric Kondo regime, this can be achieved by using $k_\text{B}T_\text{K}=wk_\text{B}T_0$, with
$T_0$ from Eq.~(\ref{eq:t0-symm}), and the Schrieffer-Wolff relation between $\rho J$ and the Anderson model parameters. One finds the result \cite{Haldane1978b,KWW1980a},
\begin{align}
  k_\text{B}T_{K} & = D_{\text{eff}}\sqrt{2\rho J}e^{-1/2\rho J  },\;\;\;\text{for $\rho J\ll 1$}, \label{eq:TK-AndersonModel-HighTdef}
\end{align}
where $D_{\text{eff}}=w\sqrt{\pi}U/4\approx 0.1829U$ is the effective half-bandwidth for an Anderson model that yields the same $T_\text{K}$ as for a Kondo model of half-bandwidth $D$, i.e.,
$D_{\text{eff}}$ plays the same role as $D$ in the prefactor in Eq.(\ref{eq:TK-KondoModel-HighTdef}). The above expression for $k_\text{B}T_\text{K}$ holds for $D\gg U$ (wide-band limit) and
in the symmetric Kondo regime of the Anderson model. It can also be applied to the asymmetric Kondo regime when both $|\varepsilon|$ and $|\varepsilon+U|$ are smaller than $D$, but
requires modification on approaching the mixed valence regime\cite{KWW1980b}.

\subsection{Kondo scale $\Gamma_\text{K}$ in d$I$/d$V$ and universal linewidths}
\label{subsec:Kondo-resonance-width}
\begin{figure*}[h!]
  \centering 
\includegraphics[width=0.98\textwidth]{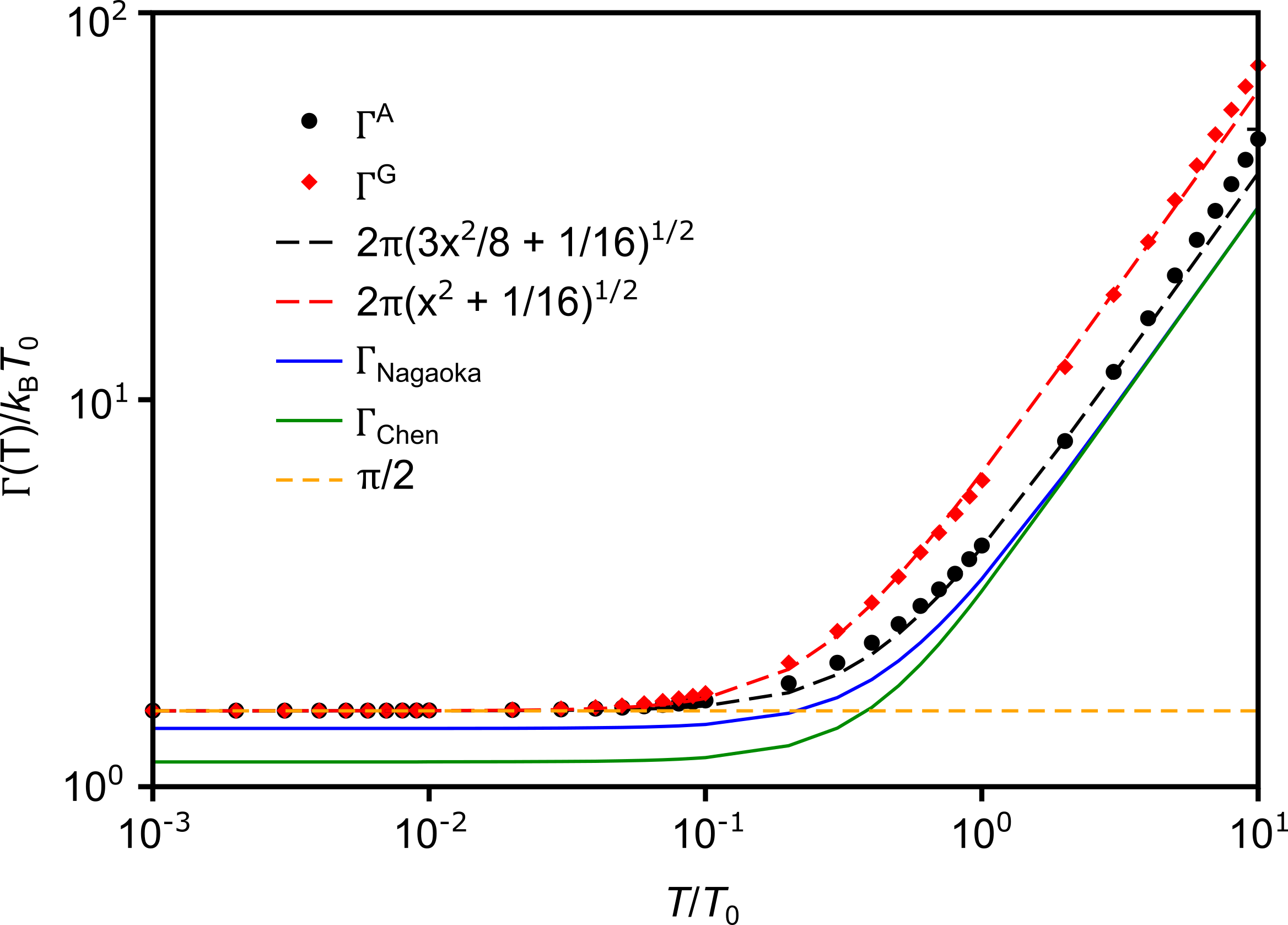}
\caption 
{\footnotesize \textbf{Temperature dependence of the HWHM of the Kondo resonance in $A(E,T)$ and d$I$/d$V$.} $\Gamma^{A,G}/k_\text{B}T_0$ with $T_0$ from Eq.~(\ref{eq:t0-symm}) are NRG calculations for $U/\gamma_0 = 25.6$. 
  Fits with square root interpolations are indicated. Also shown are the expressions $\Gamma_{\text{Nagaoka}}$ from Eq.~(\ref{eq:Nagaoka})  and $\Gamma_{\text{Chen}}$ from Eq.~(\ref{eq:Chen}). The zero temperature limit of the NRG data gives a Kondo resonance HWHM of $\pi k_\text{B} T_0/2$. The NRG calculations used $\Lambda=5$, $N_z=192$ and retained 1400 states per iteration.
}
\label{fig:Gamma}
\end{figure*}

In Fig.~\ref{fig:Gamma}, we show 
the universal curves for $\Gamma^A(T)/k_\text{B}T_0$ and $\Gamma^G(T)/k_\text{B}T_0$ versus $T/T_0$ for the symmetric Anderson model in the Kondo regime with $U/\gamma_0 = 25.6$.  For 
$\Gamma^A(T)$, we also show analytic results using Fermi liquid theory for the Anderson model by Nagaoka \textit{et al.}~\cite{Nagaoka2002} and refinements of this by Chen \textit{et al.}~\cite{Chen2021}.
Both the NRG and the Fermi liquid approaches use the strong coupling scale $T_0$, allowing them to be compared.

The NRG gives $\Gamma_\text{K}=\Gamma^G(T=0)=\Gamma^A(T=0)=1.5706 k_\text{B} T_0\approx \pi k_\text{B} T_0/2$, \textit{i.e.},
the zero temperature width of the Kondo resonance clearly approaches $\pi k_\text{B}T_0$. Previous NRG estimates of $\Gamma_\text{K}$, for a smaller $U/\gamma_0\approx 7.9$, gave $2\Gamma_\text{K}=3.0k_\text{B}T_0$ \cite{Hsu2022}. The Frota lineshape \cite{Frota1992}, an interpolation of NRG data for the zero temperature spectral function, gave $2\Gamma_\text{K}\approx 2.98k_\text{B}T_0$, consistent with our estimate. 
The Fermi liquid approaches underestimate $\Gamma_\text{K}$: the expression of Nagaoka \textit{et al.} gives $\sqrt{2}k_\text{B}T_0 = 1.41k_\text{B}T_0$ while the expression of Chen \textit{et al.} gives $\Gamma_\text{K}=1.16k_\text{B}T_0$.
The reason for this is well known: the Kondo resonance lineshape is not a Lorentzian, except for $|E|\ll k_\text{B}T_0$. Instead, it has logarithmic tails which enhance its linewidth \cite{Rosch2003}.
Estimating its linewidth requires a method like NRG which can access the crossover energy scale $E\sim \Gamma_\text{K}\sim k_\text{B}T_0$ which is outside the regime where Fermi liquid theory applies.
Despite this, the Fermi liquid approaches, which give simple expressions for $\Gamma^A(T)$ and its dimensionless form $f^A(x=T/T_0)=\Gamma^A(T)/k_\text{B}T_0$, capture the qualitative aspects of the temperature dependence. The reason for this is that they can be regarded as giving a lower bound on the line width since a Lorentzian is always narrower than the true Kondo resonance.
The Nagaoka expressions for $\Gamma^A(T)=\Gamma_{\text{Nagaoka}}^A(T)$  and $f^A_{\text{Nagaoka}}(x=T/T_0)$ read
\begin{align}
  \Gamma_{\text{Nagaoka}}^A(T) &= \sqrt{(\pi k_\text{B}T)^2+ 2(k_\text{B}T_0)^2},\label{eq:Nagaoka}\\
  f^A_{\text{Nagaoka}} (x) & = \sqrt{\pi^2x^2 + 2},\label{eq:f-a-Nagaoka}
\end{align}
while the corresponding expressions of Chen et al. read,
\begin{align}
  \Gamma_{\text{Chen}}^A(T) &= \frac{4\sqrt{2}}{\pi} \left[  \sqrt{  (k_\text{B}T_0)^4+ \left[ (k_\text{B}T_0)^2+ \frac{\pi^4}{32} (k_\text{B}T)^2 \right]^2} -(k_\text{B}T_0)^2 \right]^{1/2},\label{eq:Chen}\\
   f_{\text{Chen}}^A(x) &= \frac{4\sqrt{2}}{\pi} \left[  \sqrt{ 1 + \left[ 1 + \frac{\pi^4}{32} x^2 \right]^2} -1 \right]^{1/2},\label{eq:f-a-Chen}
\end{align}
An attempt to fit the NRG results for $\Gamma^{A,G}(T)$ in the range $10^{-3}\le T/T_0 \le 10$ with a square root expression of the form
$\Gamma^{A,G}_{\text{NRG}} = (1/2)\sqrt{(\alpha^{A,G} k_\text{B}T)^2 +(\beta^{A,G} k_\text{B}T_0)^2}$ yields for the specified range,
\begin{align}
\Gamma_{\text{NRG}}^A(T) & = \sqrt{\frac{3}{2}(\pi k_\text{B}T)^2+ (\frac{\pi k_\text{B}T_0}{2})^2},\label{eq:NRG-A}\\
f^A_{\text{NRG}} (x=T/T_0) &= \frac{\Gamma^A_{\text{NRG}}(T)}{k_\text{B}T_0} = 2\pi\sqrt{\frac{3}{8}x^2 + \frac{1}{16}},\label{eq:f-a-nrg}\\
\Gamma_{\text{NRG}}^G(T) & = \sqrt{(2\pi k_\text{B}T)^2+ (\frac{\pi k_\text{B}T_0}{2})^2},\label{eq:NRG-G}\\
f^G_{\text{NRG}} (x=T/T_0) &= \frac{\Gamma^G_{\text{NRG}}(T)}{k_\text{B}T_0} = 2\pi\sqrt{x^2 + \frac{1}{16}},\label{eq:f-g-nrg}
\end{align}
where the NRG coefficients $\beta^{A}=\beta^{G}=\pi$ give the correct width at $T=0$, i.e., $f^{A,G}_{\text{NRG}}(x=0)=\pi/2$ so $2\Gamma_\text{K}=\pi k_\text{B} T_0$. The coefficients $\alpha^{A}=\sqrt{6}\pi\approx 1.2 \times 2\pi$ and $\alpha^G=4\pi$ give reasonable fits to the NRG data in the range $10^{-3}\le T/T_0 \le 10$. It should be emphasized that these coefficients are not chosen to reproduce exactly the NRG data in the limits $x\to 0$ (low temperature limit) or $x\gg 1$ (high temperature limit) exactly, but only to obtain a good overall fit in the above range. 
Indeed,  outside the above range, i.e., for $x=T/T_0> 10$, the approximate functions $f^{A,G}_{\text{NRG}}(x)$ actually deviate significantly from the NRG data (as can already be seen in Fig.~\ref{fig:Gamma} for $x=10$). 
Even in the range $10^{-3}\leq x \le 10^1$ where $f^{A,G}_{\text{NRG}}(x)$ fit the NRG data reasonably well, there are still significant deviations, e.g., in the range $0.1\leq x \leq 0.7$ for $f^{G}_{\text{NRG}}(x)$. 
The fact that the expressions $f_{\text{NRG}}^{A,G}$ do not fit the NRG data very well, shows that the true scaling functions $\Gamma_{\text{NRG}}^{A,G}$ are indeed more complicated than a square root function.

\newpage

\section*{Supplementary Note 13: NRG spectral function and d$I$/d$V$}
    \addcontentsline{toc}{section}{Supplementary Note 13: NRG spectral function and d$I$/d$V$}
\label{sec:nrg-spectra}   
\begin{figure*}[h!]
  \centering 
\includegraphics[width=0.98\textwidth]{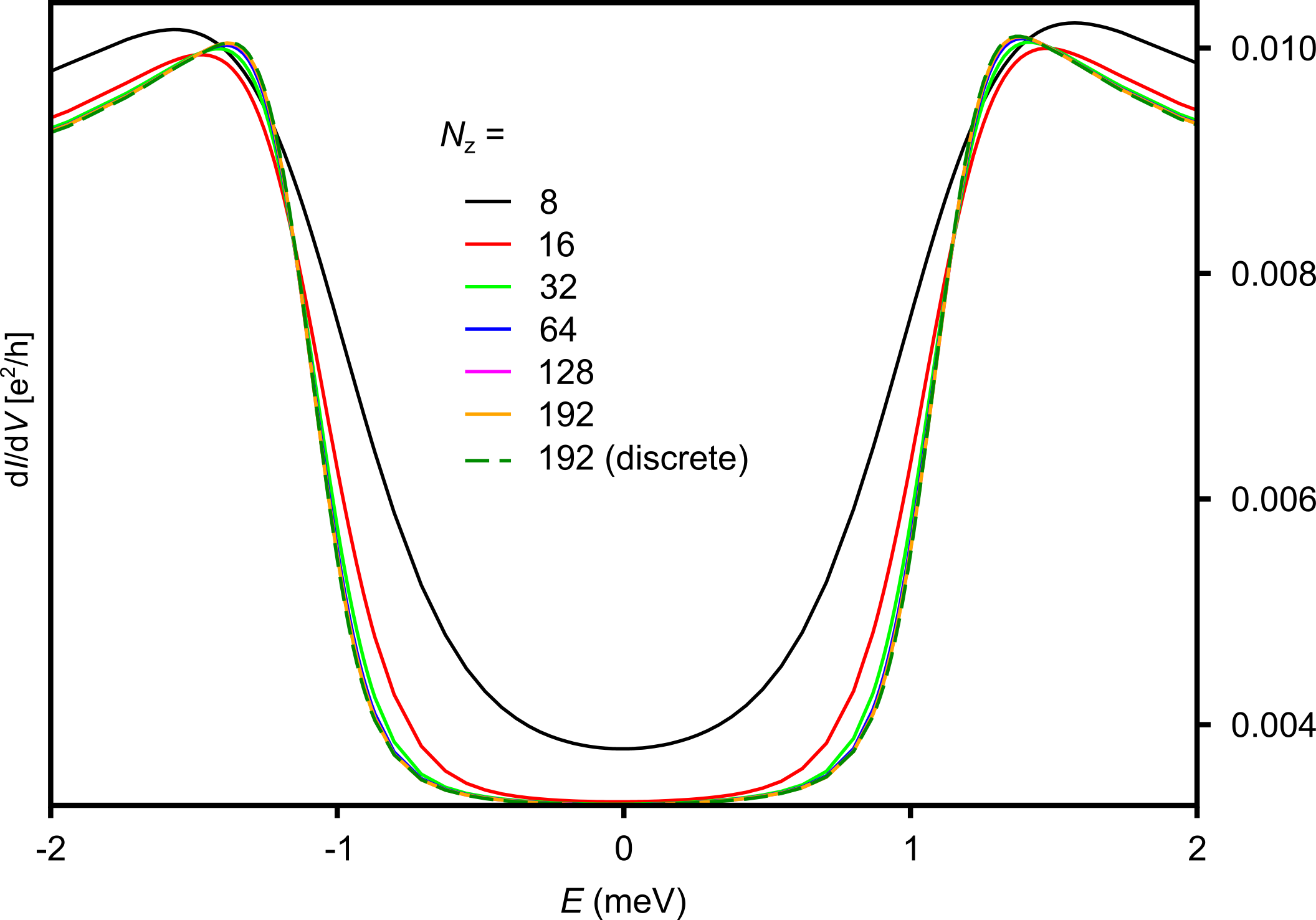}
\caption{\footnotesize \textbf{Convergence of d$I$/d$V$ spectra with $N_z$}. Illustrates how the correct thermal broadening of the inelastic steps at $E=\pm B$ in d$I$/d$V$ is captured with increasing number of bath realizations $N_z$ in the z-averaging approach to spectral functions. Convergence to the numerically exact form  (green dashed line) in Eq.~(\ref{eq:NRG-dIdV-discrete}), i.e., without broadening of the delta peaks, 
  is achieved for $N_z\gtrsim 128$. The parameters used for the Anderson model calculation are for the symmetric MTB  ($U=\SI{100}{\meV}$, $\varepsilon=\SI{-51}{\meV}$ and $\gamma =\SI{9.35}{\meV}$) at the largest 
  experimental field $B=\SI{8}{\tesla}$ and for $T=\SI{0.7}{\K}$. The NRG calculations here, and elsewhere, used $\Lambda=5$ and retained 1400 states per iteration. }
\label{fig:nz-dep}
\end{figure*}
 Due to the use of a logarithmically discretized conduction band, the NRG spectral function takes on a discrete form $A(E,T)=\sum_{m}w_m\delta(E-\varepsilon_m)$ where $\varepsilon_m$ are the many body excitations and $w_m$ are weights containing matrix elements and Boltzmann factors (or within the full density matrix approach \cite{Peters2006,Weichselbaum2007,Kugler2022}, that we use, reduced density matrix elements). In principle, such a discrete spectral function can be directly substituted into Eq.~(\ref{eq:NRG-dIdV}) and evaluated to give 
\begin{align}
\frac{dI}{dV}(eV) & = \sum_{m}\frac{w_m}{4k_\text{B}T \cosh^2((\varepsilon_m-eV)/2k_\text{B}T)}.\label{eq:NRG-dIdV-discrete}
\end{align}
Since this approach just uses the raw NRG excitations and matrix elements, it is in principle exact and includes in d$I$/d$V$ just the required thermal broadening. However, in practice, due to the 
lower logarithmic resolution of the NRG excitations $\varepsilon_m$ at higher energies, this approach only works for $|eV| \lesssim 100 k_\text{B}T$.  This would suffice to obtain the low energy (Kondo) features 
of d$I$/d$V$ in the experiment, however, an important aspect of the present  d$I$/d$V$ measurements is that they also probe the high energy Hubbard peak excitations at $\varepsilon$ and $\varepsilon+U$
of the Anderson model. In practice, therefore, we proceed by first obtaining a smooth spectral function $A(E,T)$, by replacing the 
  delta functions appearing in the discrete $A(E,T)$ by, typically, Gaussians, logarithmic Gaussians or other functions with appropriate widths. This NRG broadened 
  $A(E,T)$ is then substituted in Eq.(\ref{eq:NRG-dIdV}) in Supplementary Note 12 to obtain d$I$/d$V$. In this procedure,  it is important to demonstrate that the NRG broadening of the discrete spectral function does not influence the 
  widths of excitations and features in d$I$/d$V$, such as, for example, the slope of the sharp inelastic peaks in d$I$/d$V$ at $E=\pm B$. This problem is overcome by using the z-averaging approach to spectral functions \cite{Campo2005} with $N_z$ realizations of the discretized conduction band, and choosing $N_z\gg 1$ sufficiently large. As illustrated in Fig.~\ref{fig:nz-dep}, 
  for  $N_z\gtrsim 128$ the NRG broadening used for $A(E,T)$, which is proportional to $1/N_z$,  no longer affects the shape of the sharp inelastic steps at $E=\pm B$ in d$I$/d$V$. As required, these are 
  then only determined by the temperature $T$, which is obtained from experiment. All NRG calculations for comparison with experiment, reported in this paper, were carried out for $N_z=192$.

	\bibliographystyle{naturemag}
	\bibliography{./library}